\documentclass[nofootinbib,amsmath,twocolumn,notitlepage,preprintnumbers]{revtex4-1}
\usepackage{multirow}
\usepackage{amssymb, esvect, amsmath, graphicx, latexsym, amsthm, slashed, eso-pic}
\usepackage{xcolor}
\usepackage{hyperref}
\usepackage{amsmath}

\newcommand{\ie}{{\it i.e.}}

\newcommand{\beq}{\begin{equation}} \newcommand{\eeq}{\end{equation}}
\newcommand{\bea}{\begin{eqnarray}} \newcommand{\eea}{\end{eqnarray}}

\def\lsim{\mathrel{\raise.3ex\hbox{$<$\kern-.75em\lower1ex\hbox{$\sim$}}}}
\def\gsim{\mathrel{\raise.3ex\hbox{$>$\kern-.75em\lower1ex\hbox{$\sim$}}}}

\newcommand{\be}{\begin{eqnarray}}
\newcommand{\ee}{\end{eqnarray}}

\newcommand{\benum}{\begin{enumerate}}
\newcommand{\eenum}{\end{enumerate}}
\newcommand{\bi}{\begin{itemize}}
\newcommand{\ei}{\end{itemize}}

\begin{document}

\preprint{FERMILAB-PUB-22-923-T}

\title{The Cherenkov Telescope Array Will Test Whether Pulsars Generate the Galactic Center Gamma-Ray Excess}

\author{Celeste Keith$^{a,b}$}
\thanks{ORCID: https://orcid.org/0000-0002-3004-0930}

\author{Dan Hooper$^{a,b,c}$}
\thanks{ORCID: http://orcid.org/0000-0001-8837-4127}

\author{Tim Linden$^{d}$}
\thanks{ORCID: http://orcid.org/0000-0001-9888-0971}

\affiliation{$^a$University of Chicago, Kavli Institute for Cosmological Physics, Chicago IL, USA}
\affiliation{$^b$University of Chicago, Department of Astronomy and Astrophysics, Chicago IL, USA}
\affiliation{$^c$Fermi National Accelerator Laboratory, Theoretical Astrophysics Group, Batavia, IL, USA}
\affiliation{$^d$Stockholm University and The Oskar Klein Centre for Cosmoparticle Physics, AlbaNova, 10691 Stockholm, Sweden}

\date{\today}

\begin{abstract}

The GeV-scale gamma-ray excess observed from the region surrounding the Galactic Center has been interpreted as either the products of annihilating dark matter particles, or as the emission from a large population of faint and centrally-located millisecond pulsars. If pulsars are responsible for this signal, they should also produce detectable levels of TeV-scale emission. In this study, we employ a template-based analysis of simulated data in an effort to assess the ability of the Cherenkov Telescope Array (CTA) to detect or constrain the presence of this emission, providing a new and powerful means of testing whether millisecond pulsars are responsible for the observed excess. We find that after even a relatively brief observation of the Inner Galaxy, CTA will be able to definitively detect this TeV-scale emission, or rule out pulsars as the source of the Galactic Center Gamma-Ray Excess.

\end{abstract}
 
\maketitle

\section{Introduction}

An excess of GeV-scale gamma rays from the region surrounding the Galactic Center has been identified in the publicly available data collected by the Fermi Gamma-Ray Space Telescope~\cite{Goodenough:2009gk,Hooper:2010mq,Hooper:2011ti}. The spectral shape, morphology, and intensity of this emission are each consistent with arising from the annihilation of dark matter particles. More specifically, the characteristics of this signal match those predicted from a $m_X \sim 50 \, {\rm GeV}$ dark matter particle with an annihilation cross section of $\sigma v \sim (1-2) \times 10^{-26} \, {\rm cm}^3/{\rm s}$, and that is distributed slightly more steeply around the Galactic Center than described by the canonical Navarro-Frenk-White profile, $\rho \propto r^{-1.2}$~\cite{Daylan:2014rsa,Calore:2014xka,Calore:2014nla,Fermi-LAT:2017opo,DiMauro:2021raz,Cholis:2021rpp}. The possibility that Fermi could be detecting dark matter annihilation products has generated a great deal of interest (see, for example, Refs.~\cite{Berlin:2014tja,Izaguirre:2014vva,Ipek:2014gua,Agrawal:2014una,Berlin:2014pya,Abdullah:2014lla,Martin:2014sxa,Alves:2014yha,Boehm:2014hva,Berlin:2015wwa,Karwin:2016tsw}).

The leading alternative explanation for this excess is that it is instead generated by a large population of unresolved millisecond pulsars~\cite{Hooper:2010mq,Abazajian:2010zy,Hooper:2013nhl,Gordon:2013vta,Cholis:2014lta,Hooper:2013nhl,Petrovic:2014xra,Brandt:2015ula,Hooper:2015jlu,Hooper:2016rap}. This possibility is motivated in large part by the fact that pulsars are the only known class of astrophysical objects that produce a gamma-ray spectrum with a shape that is similar to that of the Galactic Center Gamma-Ray Excess. As the stellar population of the Galactic Bulge consists predominantly of old stars, we expect relatively few young or middle-aged pulsars to be found in the Inner Galaxy. In contract, pulsars with millisecond-scale periods (known as millisecond pulsars, or recycled pulsars) can remain bright for billions of years, and could plausibly be more abundant in this region of the Galaxy.

The hypothesis that millisecond pulsars produce the observed gamma-ray excess was elevated substantially in 2015, when two independent groups claimed to have identified evidence that the photons constituting this excess are spatially clustered~\cite{Lee:2015fea,Bartels:2015aea}. Furthermore, it was claimed in Refs.~\cite{Bartels:2017vsx,Macias:2016nev,Macias:2019omb,Pohl:2022nnd} that this excess is not distributed with spherical symmetry around the Galactic Center (as would be expected from dark matter), but is instead correlated with the distribution of stars that make up the Galactic Bulge and Bar. If true, these results would have significantly favored the conclusion that the excess originates from a population of near-threshold astrophysical point-sources, rather than from annihilating dark matter. 

The conclusions described in the previous paragraph have been strongly challenged in recent years (for a summary, see Ref.~\cite{Hooper:2022bec}). In particular, the small scale power identified in Ref.~\cite{Lee:2015fea} was shown in Refs.~\cite{Leane:2019xiy,Leane:2020pfc,Leane:2020nmi} to be an artifact of insufficiently understood backgrounds, causing smooth signals to be mischaracterized as clumpy. Furthermore, the evidence for unresolved point sources claimed in Ref.~\cite{Bartels:2015aea} was found to disappear when the point source catalog was updated~\cite{Zhong:2019ycb}. Even more recently, it was shown in Refs.~\cite{DiMauro:2021raz} and~\cite{Cholis:2021rpp} that the angular distribution of the excess emission is best fit by a spherical and dark matter-like morphology, and does not significantly correlate with any known stellar populations; the Fermi data prefers the excess to have a bulge-like morphology only when the background model provides a poor fit to the overall dataset~\cite{McDermott:2022zmq}. To the best of our ability to measure, the gamma-ray excess appears to be both smooth, and distributed with spherical symmetry around the Galactic Center. Furthermore, Fermi data has been used to place strong constraints on the luminosity function of any point source population that might be responsible for this signal~\cite{Bartels:2017xba,Zhong:2019ycb,Dinsmore:2021nip}, posing a significant challenge for millisecond pulsar interpretations of the gamma-ray excess.

None of the results described above preclude the possibility that the gamma-ray excess could be produced by a large population of very faint millisecond pulsars, distributed with approximate spherical symmetry around the Galactic Center. If this is the case, however, there are accompanying signals that should be present in the GeV~\cite{Bartels:2017xba,Dinsmore:2021nip}, radio~\cite{Calore:2015bsx}, and X-ray~\cite{Haggard:2017lyq} bands. Such signals could be used to potentially constrain or confirm the hypothesis that millisecond pulsars generate the gamma-ray excess. Furthermore, observations by the High Altitude Water Cherenkov (HAWC) Observatory and the Large High Altitude Air Shower Observatory (LHAASO) have shown that young and middle-aged pulsars are typically surrounded by bright, spatially-extended, multi-TeV emitting regions known as ``TeV halos''~\cite{Hooper:2017gtd,Linden:2017vvb,HAWC:2021dtl,HAWC:2019tcx}. Even more recently, it has been shown (at the 99\% C.L.) that millisecond pulsars also generate TeV halos~\cite{Hooper:2021kyp}. If this result is robustly confirmed, it will be possible to use ground-based gamma-ray telescopes, such as the Cherenkov Telescope Array (CTA), to search for the TeV-scale emission from any millisecond pulsars that might be present near the Galactic Center. Such a measurement would provide an independent -- and potentially definitive -- measurement of our Inner Galaxy's millisecond pulsar population~\cite{Hooper:2018fih}. 

In this paper, we perform a template-based analysis of simulated data to assess the ability of the Cherenkov Telescope Array (CTA) to identify and measure the very high-energy gamma-ray emission from a population of millisecond pulsars that are located in the region surrounding the Galactic Center. When the intensity of this emission is normalized to the measurements of other pulsars by HAWC and LHAASO, we find that if the Galactic Center Gamma-Ray Excess originates from pulsars, this source population should produce the dominant contribution to the TeV-scale emission from the Inner Galaxy. Such a signal would be easily detectable by CTA, even after a relatively short observation of this region. We conclude that CTA will either be able to clearly identify the TeV halo emission associated with such a pulsar population, or rule out the hypothesis that the Galactic Center Gamma-Ray Excess originates from millisecond pulsars.

\section{Gamma-Ray Emission From TeV Halos Around Millisecond Pulsars}
\label{sec:sec2}

%In this section, we will describe the origin of TeV scale halos and the calculation of the full gamma-ray spectrum produced by our TeV scale pulsars via inverse Compton scattering of high energy electrons and positrons, as described in~\cite{Sudoh:2021avj}.

In 2017, observations by the HAWC Observatory identified bright, multi-TeV emission from the regions surrounding the nearby Geminga and Monogem pulsars~\cite{HAWC:2020hrt,Abeysekara:2017hyn,Abeysekara_2017_2} (for earlier observations by Milagro, see Ref.~\cite{Abdo:2009ku}). It has since been found that similar TeV halos are present around most, if not all, young and middle-aged pulsars~\cite{Linden:2017vvb,HAWC:2021dtl,Sudoh:2021avj}. This emission is produced through the inverse Compton scattering of very high-energy electrons and positrons, and the intensity of the observed emission implies that $\mathcal{O}(10\%)$ of the pulsars' total energy budget (i.e., spindown power) goes into the acceleration of such particles~\cite{Hooper:2017gtd,Sudoh:2021avj}. Applying these characteristics to the larger population of pulsars, one expects TeV halos to dominate the Milky Way's diffuse gamma-ray emission at TeV-scale energies~\cite{Linden_2018}, and to contribute significantly to the isotropic gamma ray background~\cite{Xu:2021ncy}. 

Until recently, it was not clear whether or not pulsars with millisecond-scale periods are also surrounded by TeV halos. Based on theoretical considerations, it was generally expected that millisecond pulsars should produce TeV-scale emission in their magnetospheres, at a level similar to that which takes place among young and middle-aged pulsars~\cite{Venter:2015gga,Bednarek:2016gpp,Venter:2015oza}. On the other hand, young and middle-aged pulsars are thought to accelerate TeV-scale electrons as they reach the termination shock, and it is not clear to what extent this might occur within millisecond pulsars~\cite{2011ApJ...741...39S,Gaensler:2006ua}. Some light was shed on these questions recently, when an analysis of publicly available HAWC data demonstrated that millisecond pulsars also generate TeV emission, with an efficiency similar to that of young and middle-aged pulsars~\cite{Hooper:2021kyp} (see also, Ref.~\cite{Hooper:2018fih}). More specifically, that study considered 37 nearby and high-spindown power millisecond pulsars, finding (at the 99\% confidence level) that these sources produce TeV-scale emission with a luminosity proportional to their spindown power. Furthermore, the efficiency of this emission was constrained to lie between 39-108\% of that measured for the TeV halo associated with the middle-aged pulsar, Geminga. Future observations by telescopes including HAWC, LHAASO, and CTA should be able to confirm and further refine this conclusion, potentially confirming that millisecond pulsars have TeV halos, and also measuring the spectra, intensity, and morphology of these sources.

In calculating the gamma-ray emission from TeV halos, we adopt the following parameterization for the spectrum of the injected electrons and positrons:
\be \label{elecspec}
\!\begin{aligned}%[b]
\frac{dN_{e}}{dE_{e}} = E^{-\alpha}_{e}e^{-E_{e}/E_{\mathrm{cut}}},
\end{aligned} 
\ee
where observations indicate that $\alpha \sim 1.5-1.9$ and $E_{\mathrm{cut}} \sim 30-100 \,{\rm TeV}$ for typical TeV halos. These particles lose energy through both inverse Compton scattering and synchrotron processes, producing emission in the gamma-ray and radio bands, respectively~\cite{RevModPhys.42.237}. These processes lead to the following energy loss rate: 

\be \label{energyloss_pulsars}
\begin{aligned} 
-\frac{dE_{e}}{dt} &= \sum_{i}\frac{4}{3}\sigma_{T}u_{i}S_{i}(E_{e})\bigg(\frac{E_{e}}{m_{e}}\bigg)^{2} + \frac{4}{3}\sigma_{T}u_{\textrm{mag}}\bigg(\frac{E_{e}}{m_{e}}\bigg)^2 \\
&\equiv b(E_e) \, \bigg(\frac{E_e}{{\rm TeV}}\bigg)^2, \nonumber 
\end{aligned}
\ee
where $\sigma_{T}$ is the Thomson cross section and 
\be \label{b}
\begin{aligned}%[b]
b &\approx 1.02 \times 10^{-13} \, \mathrm{ TeV/s} \\
&\times \bigg(\sum_{i}\frac{u_{i}}{\mathrm{eV/cm^3}}S_i(E_e)+\frac{u_{\mathrm{mag}}}{\mathrm{eV/cm^3}}\bigg).
\end{aligned} 
\ee

The sum in these expressions is carried out over the relevant components of the radiation field. In our analysis, we adopt a three component radiation model, consisting of the cosmic microwave background (CMB), infrared emission, and starlight. We treat the spectra of these radiation components as blackbodies with temperatures given by T$_{\mathrm{CMB}}$ = 2.7 K, T$_{\mathrm{IR}}$ = 20 K, and T$_{\mathrm{star}}$ = 5000 K. In normalizing the infrared and starlight components, we adopt two models, with energy densities equal to $\rho_{\rm IR}= \rho_{\rm star} = 3 \, {\rm eV/cm}^3 \,({\rm ``max"})$ and $\rho_{\rm IR}= \rho_{\rm star} =0.6 \, {\rm eV/cm}^3 \,({\rm ``min"})$. For the energy density of the magnetic field, we adopt $u_{\mathrm{mag}} = 0.224 \, {\rm eV/cm}^3$ (corresponding to $B=3 \, \mu {\rm G})$ and $u_{\mathrm{mag}} = 2.5 \, {\rm eV/cm}^3 (B=10 \, \mu {\rm G})$ in the max and min models, respectively.\footnote{Note that our ``max'' and ``min'' models for the radiation and magnetic fields are defined such that they maximize or minimize the resulting gamma-ray emission, respectively.}

At very high energies, inverse Compton scattering takes places in the Klein-Nishina regime, leading to approximately the following degree of suppression~\cite{Schlickeiser_2010}:
\be \label{KN}
\begin{aligned}%[b]
S_{i}(E_{e})\approx \frac{45m^{2}_{e}/64\pi^2 T^2_{i}}{(45 m^2_e/64\pi^2 T^{2}_{i})+(E_e^{2}/m^2_e)}.
\end{aligned} 
\ee

The instantaneous spectrum of inverse Compton emission from an electron of energy, $E_e$, is given by
\be \label{specIC}
\begin{aligned}%[b]
\frac{dN_{\gamma}}{dE_{\gamma}}(E_{\gamma}, E_{e}) \propto \int \frac{dn}{d\epsilon}(\epsilon)\frac{d\sigma_{ICS}}{dE_{\gamma}}(\epsilon, E_{\gamma}, E_{e}) d\epsilon,
\end{aligned} 
\ee
where $dn/d\epsilon$ is the spectrum of the target radiation, and the differential cross section is given by~\cite{Aharonian:1981spy}
\begin{eqnarray}
\frac{d\sigma_{ICS}}{dE_{\gamma}}(\epsilon,E_{\gamma},E_{e})&=&\frac{3\sigma_{T}m_{e}^{2}}{4\epsilon E_{e}^{2}}\,
\bigg[ 1+\bigg( \frac{z^{2}}{2(1-z)}\bigg)  \\
+\bigg(\frac{z}{\beta (1-z)}\bigg)&-&\bigg(\frac{2z^{2}}{\beta^{2}(1-z)}\bigg)
-\bigg(\frac{z^{3}}{2\beta (1-z)^{2}}\bigg) \nonumber \\
&-&\bigg(\frac{2z}{\beta (1-z)}\bigg)\ln\bigg(\frac{\beta (1-z)}{z}\bigg) \bigg], \nonumber
\end{eqnarray}
where $z \equiv E_{\gamma}/E_{e}$ and $\beta \equiv 4\epsilon E_{e}/m_{e}^{2}$.

\begin{figure}
\includegraphics[width=3.2in,angle=0]{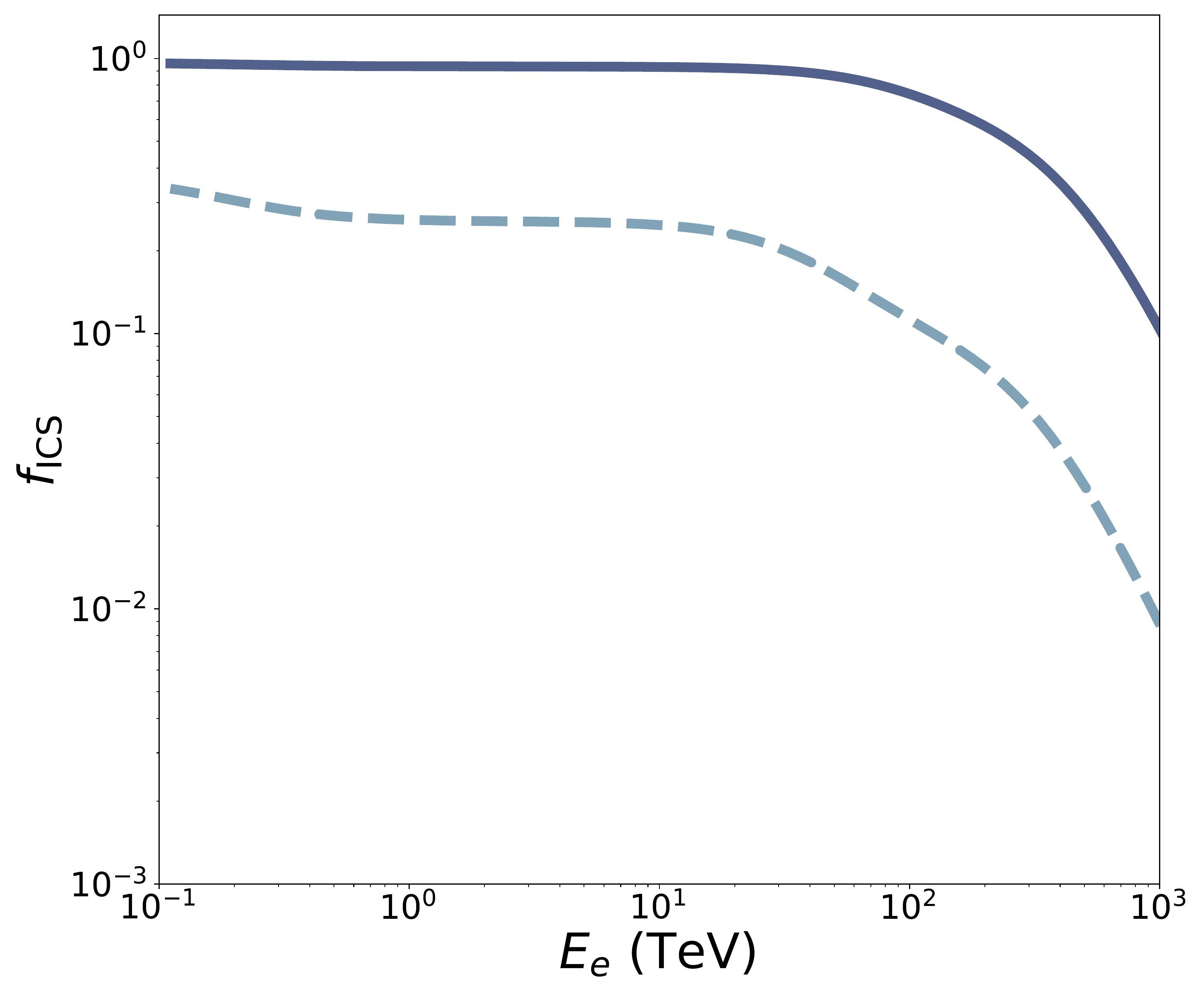}
\caption{The fraction of the instantaneous electron energy loss rate that goes into inverse Compton scattering (as opposed to synchrotron) in our ``max'' and ``min'' models for the Inner Galaxy's radiation and magnetic fields.}
\label{fICS}
\end{figure}

Since we are interested in the steady state inverse Compton emission produced from the Inner Galaxy's millisecond pulsar population, we can integrate the spectrum described by Eq.~\ref{specIC} over the lifetime of an electron to obtain the total emission from a single injected electron. In performing this calculation, we take into account the energy that is lost to synchrotron, as shown in Fig.~\ref{fICS}. We then integrate this result over the spectrum of injected electrons, as parameterized in Eq.~\ref{elecspec}, to obtain the spectrum of inverse Compton emission that is generated per unit energy injected from millisecond pulsars in the form of very high-energy electrons and positrons. 

From this spectrum of the inverse Compton emission per unit of leptonic power, $dN_{\gamma}/dE_{\gamma}dL_e$, we can calculate the spectrum and angular distribution of the gamma-ray emission from the Inner Galaxy's millisecond pulsar population:
\begin{align} 
\label{jfactor}
\frac{dN_{\gamma}}{dE_{\gamma}}(E_{\gamma}, \Delta \Omega) &= \frac{1}{4\pi} \int_{\Delta \Omega} \int_{los} \frac{dN_{\gamma}}{dE_{\gamma} dL_e}  \,  \eta \dot{E}  \, n_{\rm MSP}(r) \,dl \, d\Omega.
\end{align}
The integrals in this expression are performed over a given angular bin, $\Delta \Omega$, and over the line-of-sight, $los$. The quantity, $\dot{E}$, is the average spindown power per millisecond pulsar, and $\eta$ is the average fraction of the spindown power that goes into the production of electrons and positrons with $E_e > 100 \, {\rm GeV}$. For the spatial distribution of millisecond pulsars, we adopt $n_{\rm MSP} \propto r^{-2.4}$, where $r$ is the distance from the Galactic Center. This profile was selected because it provides provides a good fit to the observed morphology of the Galactic Center Gamma-Ray Excess~\cite{Daylan:2014rsa,Calore:2014xka,Calore:2014nla,Fermi-LAT:2017opo,DiMauro:2021raz,Cholis:2021rpp}.

\begin{figure*}
\includegraphics[width=7.5in,angle=0]{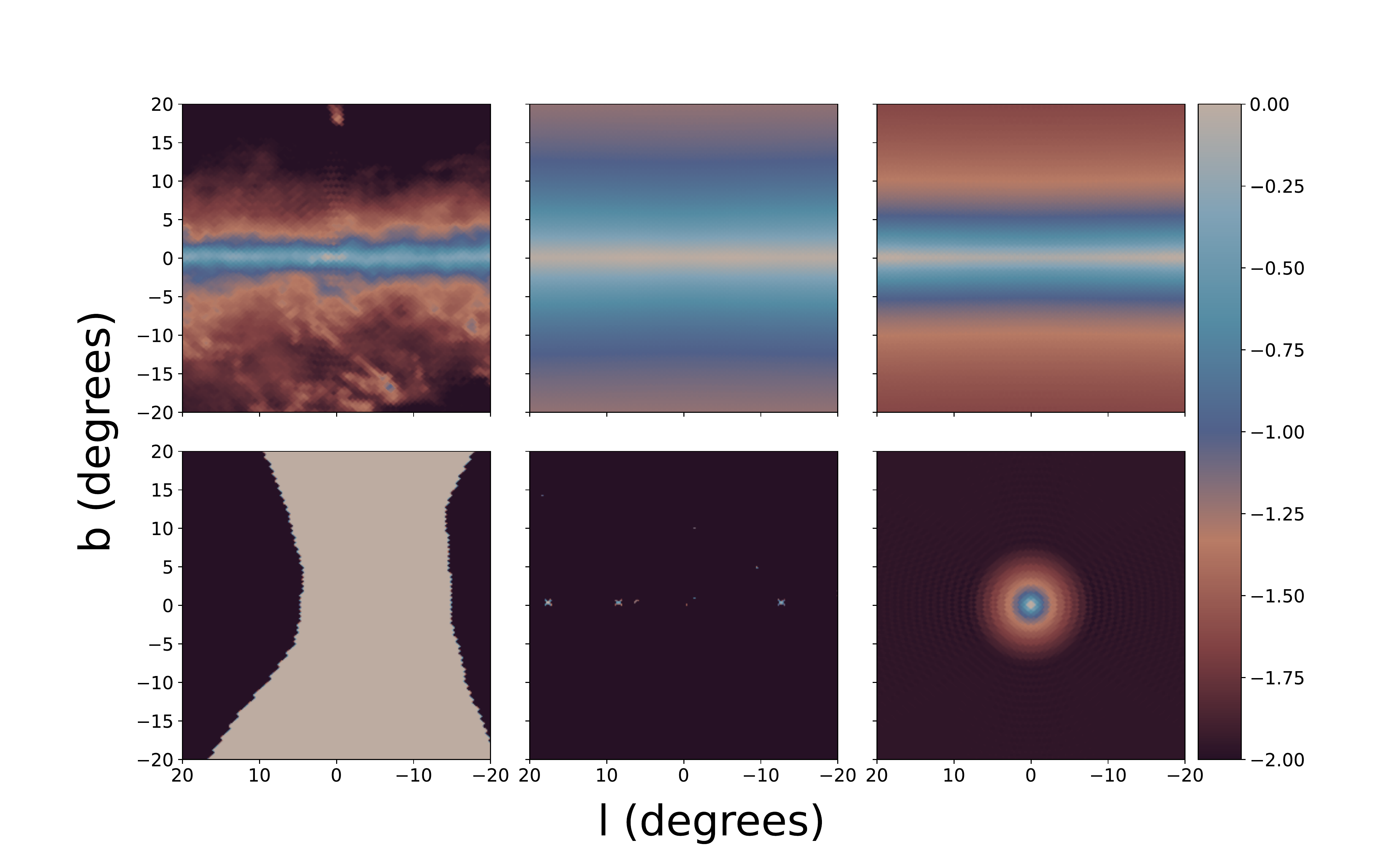}
\caption{The spatial templates used in our analysis, as evaluated at 0.1 TeV. In the top row, these templates correspond (from left-to-right) to the emission from pion production, inverse Compton scattering, and bremsstrahlung, as generated using the publicly available code GALPROP~\cite{2011CoPhC.182.1156V,2005ApJ...622..759G}. In the bottom row, we show the templates associated with the Fermi bubbles~\cite{Fermi-LAT:2014sfa}, the point sources described by the Fermi 4FGL-DR2 catalog~\cite{Fermi-LAT:2019yla}, and for the emission from the TeV halos associated with millisecond pulsars. The scale is logarithmic, and the brightest point in each frame is normalized to unity. Note that the region-of-interest used in our analysis is a $3^{\circ}$ radius circle centered at the location of the Galactic Center.}
\label{alltemplates}
\end{figure*}

The gamma-ray and radio emission from millisecond pulsars is powered by the loss of rotational kinetic energy, and thus the energy budget of such an object is set by its spindown power, $\dot{E} = 4 \pi^2 I (dP/dt)/P^3$, where $I$ and $P$ are the pulsar's moment of inertia and period. In Ref.~\cite{Hooper:2018fih}, the contents of the Fermi pulsar catalog~\cite{Fermi-LAT:2013svs} were used to estimate the efficiency at which millisecond pulsars produce GeV-scale emission, finding that such objects produce an average luminosity in the GeV-band (\ie, integrated above 0.1 GeV) equal to $\langle \eta_{\rm GeV} \rangle \approx 0.12$ times their spindown power. This quantity, however, does not take into account the fact that some pulsars have beams which are not aligned in our direction. Combining these factors, the luminosity of the GeV-scale emission from the Inner Galaxy's millisecond pulsar population can be expressed as follows:
\begin{align}
L_{\mathrm{GeV}} = \langle \eta_{\rm GeV} \rangle \, f_{\mathrm{beam}} \dot{E}_{\mathrm{tot}},
\end{align}
where $f_{\rm beam}$ is the fraction of millisecond pulsars whose gamma-ray beam is pointed in the direction of Earth, and $\dot{E}_{\rm tot}$ is the sum of the spindown power of all of the millisecond pulsars in a given region of the Inner Galaxy. Comparing this to the intensity of the Galactic Center Gamma-Ray Excess observed within a $0.5^{\circ}$ radius around the Galactic Center, $L_{\rm GCE} \approx 2 \times 10^{36} \, {\rm erg/s}$ ($>0.1 \, {\rm GeV}$), we can determine the total spindown power of the pulsar population required to accommodate $L_{\rm GeV}=L_{\rm GCE}$, as a function of $\langle \eta_{\rm GeV}\rangle$ and $f_{\rm beam}$. From this total spindown power and the TeV halo efficiency, $\eta$, the intensity of the TeV halo emission from the Inner Galaxy's millisecond pulsar population will be equal to
\begin{align}
L_{\rm TeV} &= \dot{E}_{\rm tot} \, \eta = \frac{L_{\rm GeV} \, \eta}{\langle \eta_{\rm GeV}\rangle f_{\rm beam}} \nonumber \\
& = f_{\rm GCE} \, \frac{L_{\rm GCE} \, \eta}{\langle \eta_{\rm GeV}\rangle f_{\rm beam}}, 
\label{fGCE}
\end{align}
where $f_{\rm GCE} \equiv L_{\rm GeV}/L_{\rm GCE}$ is the fraction of the Galactic Center Gamma-Ray Excess that is generated by millisecond pulsars.

\begin{figure*}
\includegraphics[width=3.25in,angle=0]{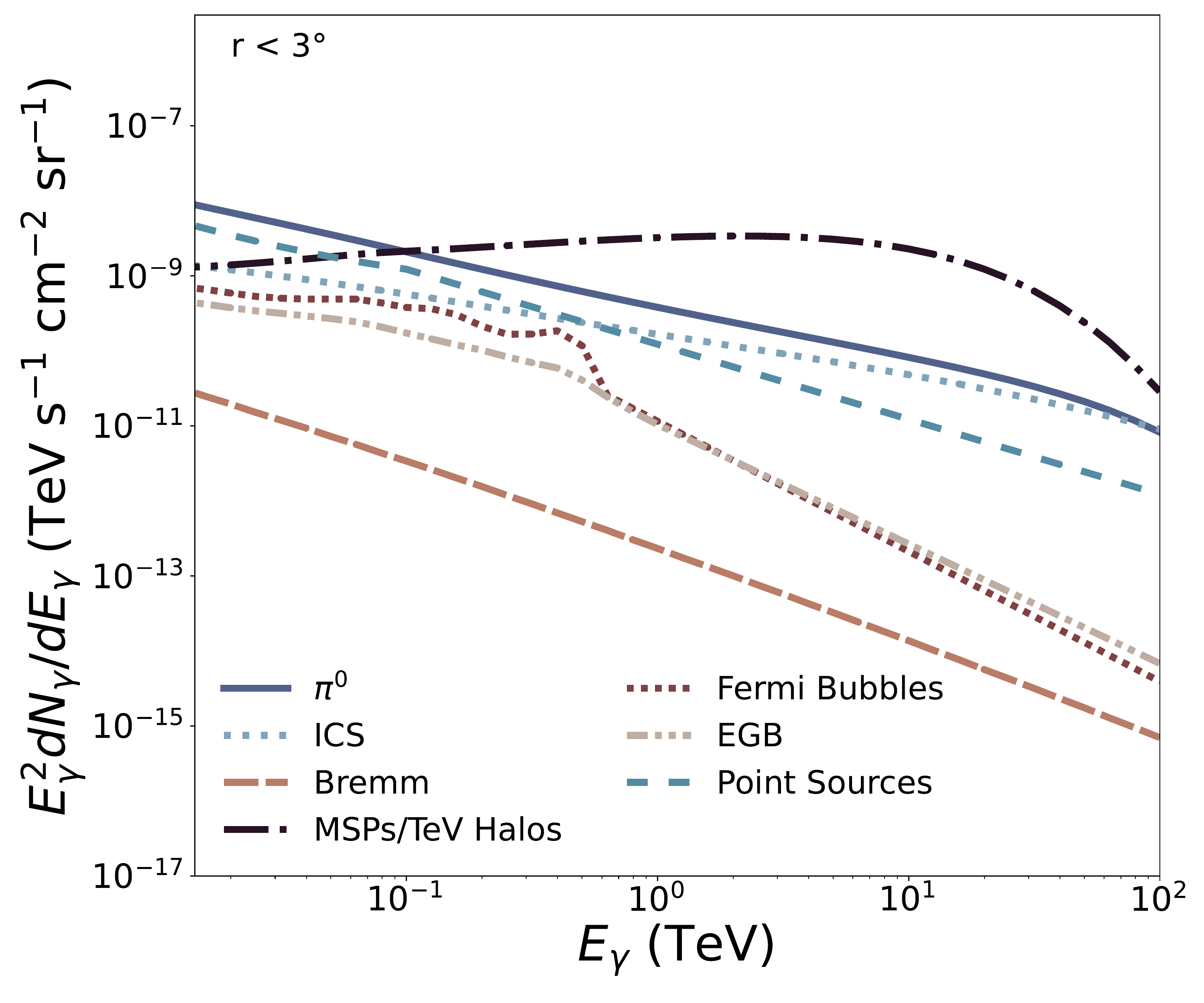}
\includegraphics[width=3.25in,angle=0]{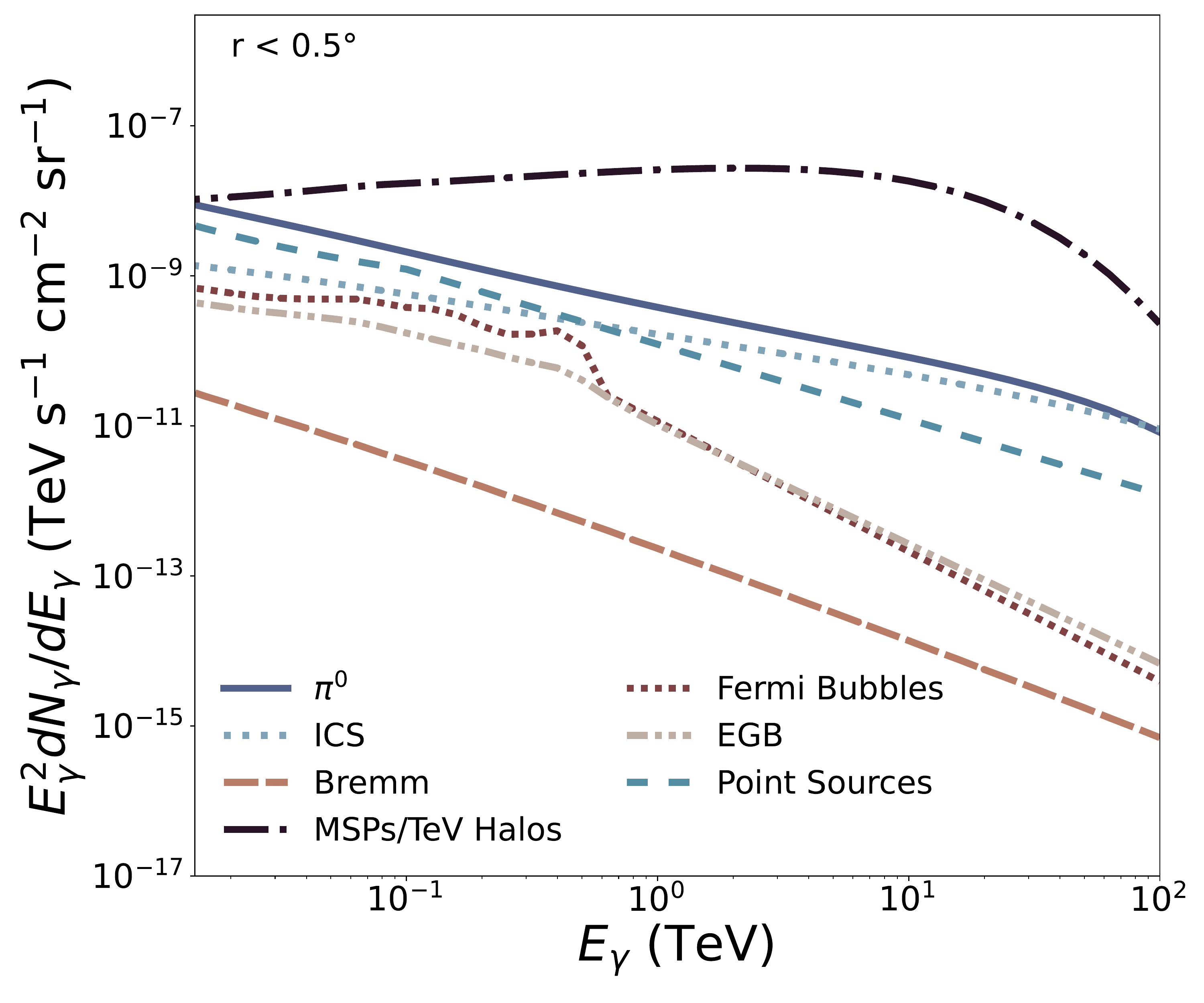}
\caption{The gamma-ray spectra of the various components of our background model, and from the TeV halos associated with millisecond pulsars in the Inner Galaxy, averaged over a $3^{\circ}$ (left) or $0.5^{\circ}$ (right) radius circle centered around the Galactic Center. For each of the three components of the Galactic diffuse emission (pion decay, inverse Compton scattering, and bremsstrahlung), we adopt the default GALPROP parameters as described in the text. In calculating the emission associated with TeV halos, we have assumed that the entire Galactic Center Gamma-Ray Excess is generated by millisecond pulsars, and have adopted $\alpha=1.5$, $E_{\rm cut}=30 \, {\rm TeV}$, $\eta =0.1$, $\langle \eta_{\rm GeV}\rangle =0.12$, $f_{\rm beam}=0.5$, and the ``min'' model for the radiation and magnetic fields. For comparison, we also show the spectra associated with the Fermi Bubbles~\cite{Fermi-LAT:2014sfa}, the extragalactic gamma-ray background (EGB)~\cite{Fermi-LAT:2014ryh}, and the point sources contained within the Fermi 4FGL-DR2 catalog~\cite{Fermi-LAT:2019yla}. If the gamma-ray excess is generated by millisecond pulsars, their TeV halos should dominate the TeV-scale emission from the Inner Galaxy.}
\label{spectra1}
\end{figure*}

\begin{figure*}
\includegraphics[width=3.25in,angle=0]{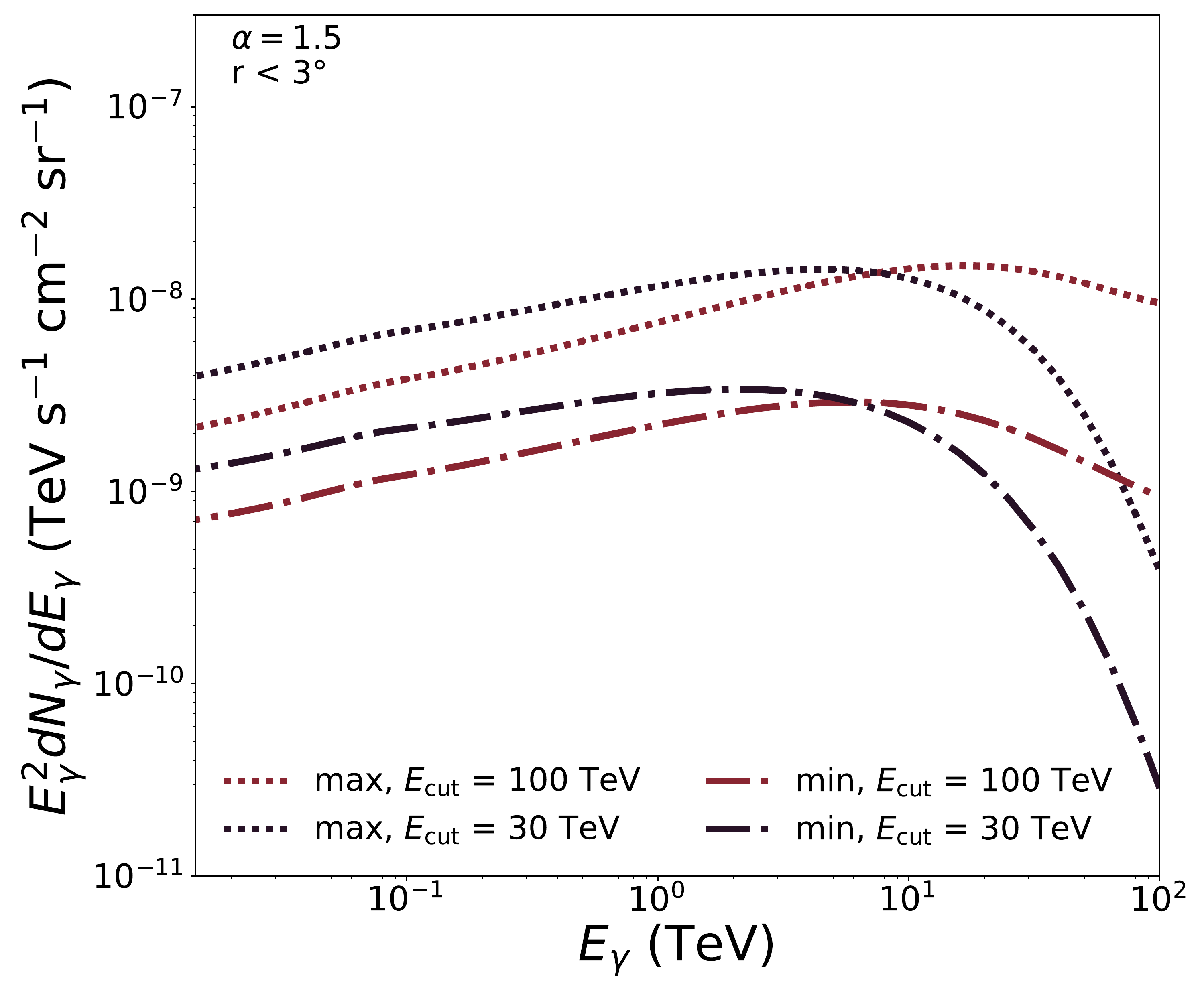}
\includegraphics[width=3.25in,angle=0]{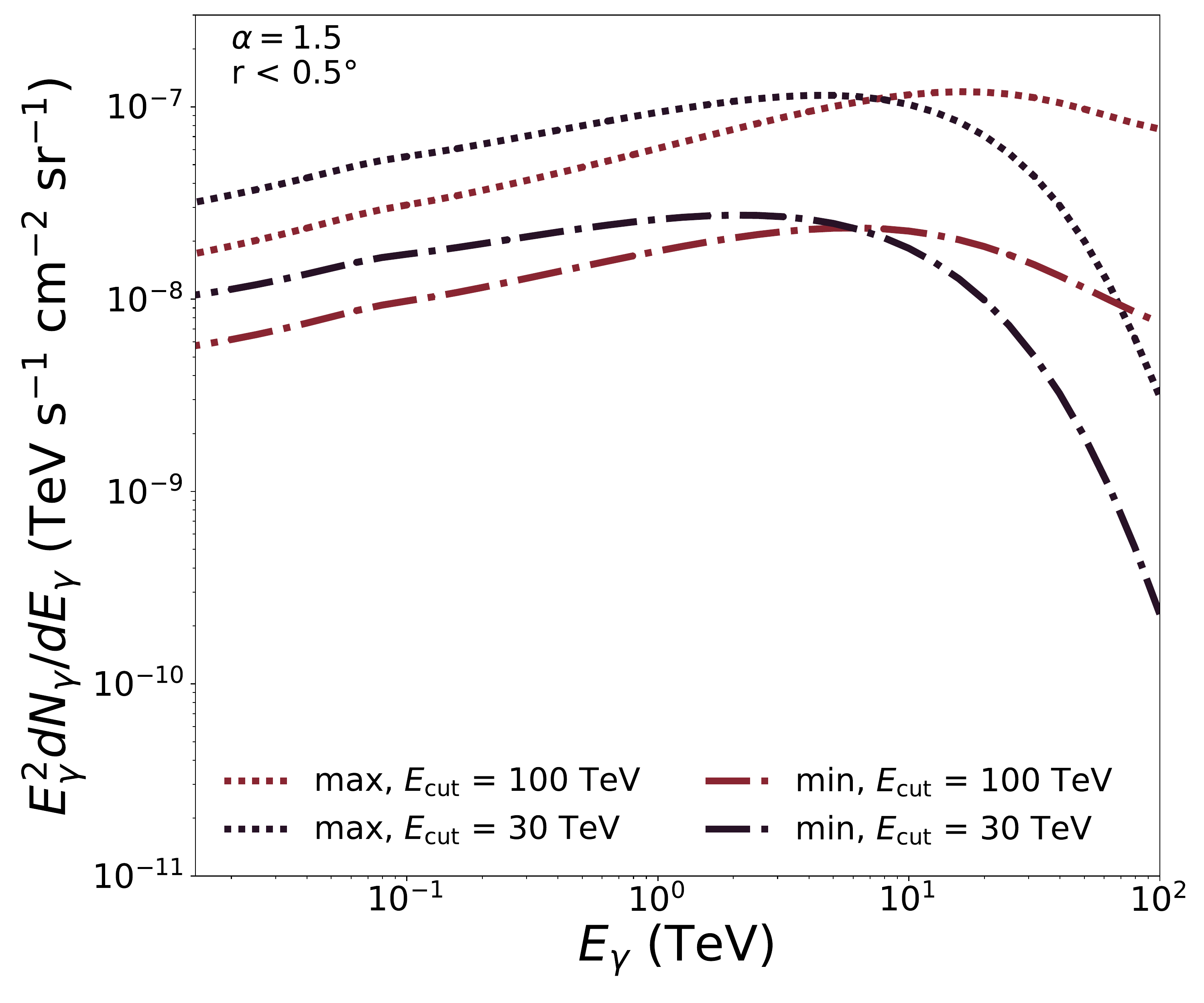} \\
\includegraphics[width=3.25in,angle=0]{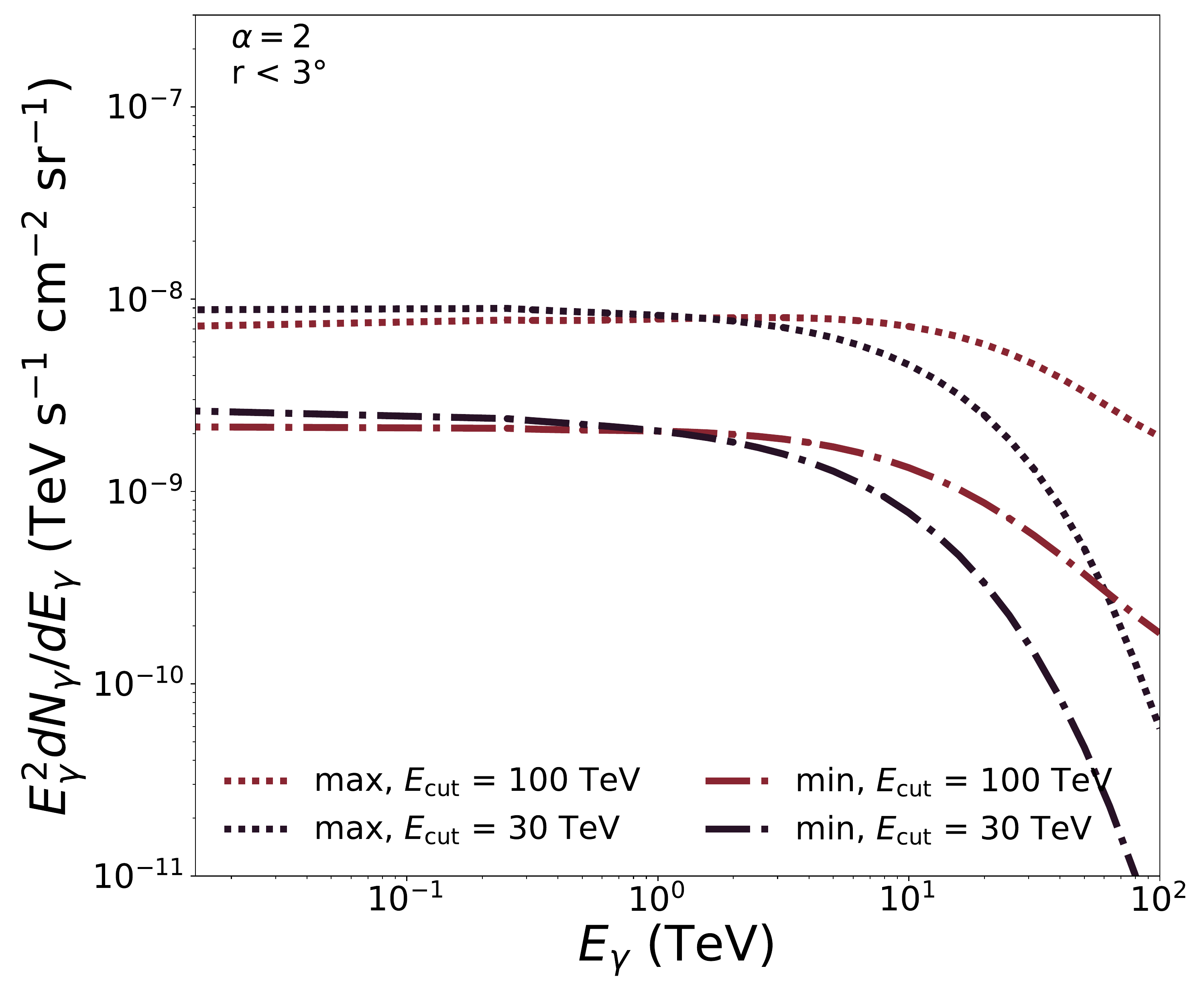}
\includegraphics[width=3.25in,angle=0]{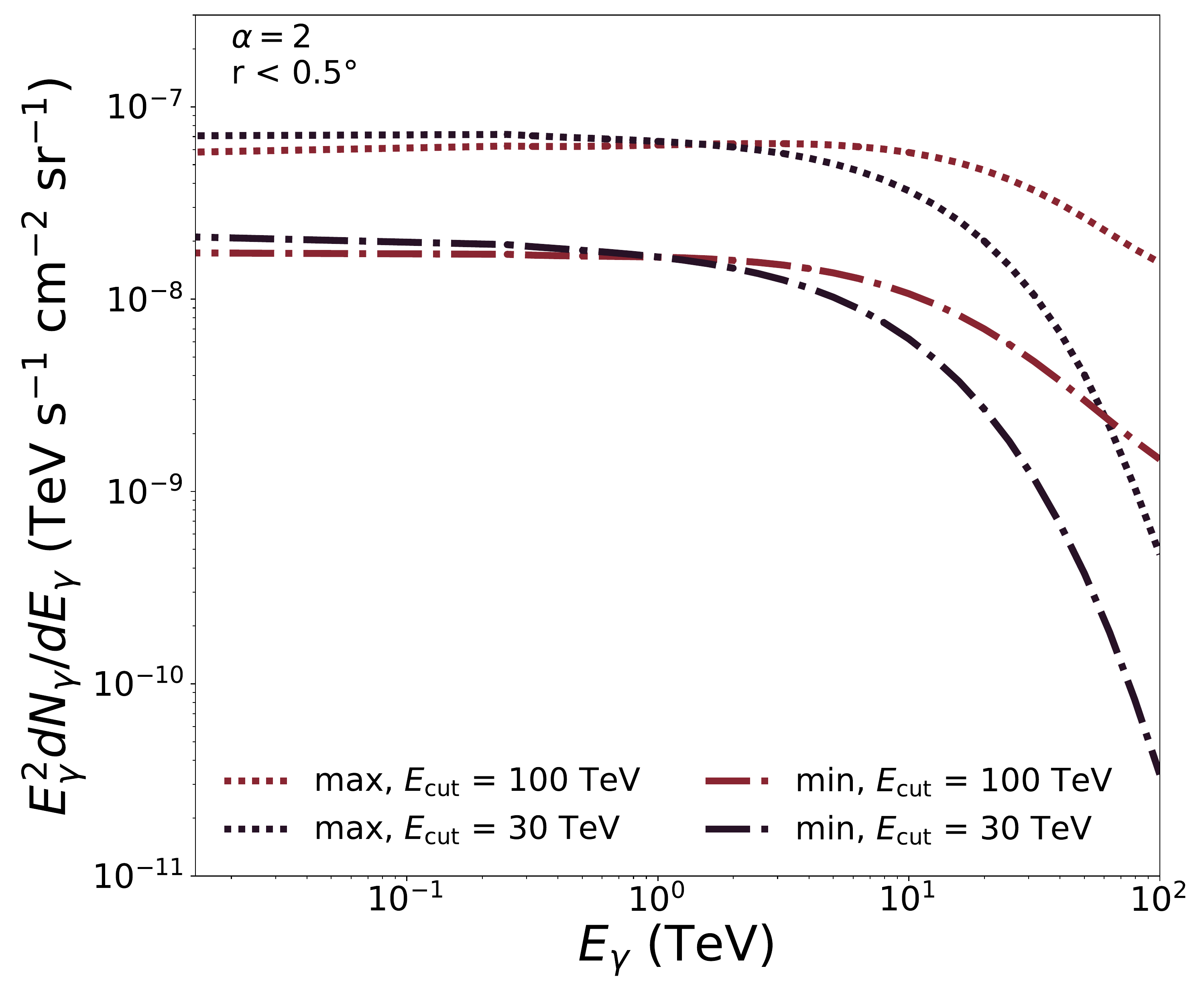}
\caption{The gamma-ray spectrum from the TeV halos associated with millisecond pulsars in the Inner Galaxy, averaged over a $3^{\circ}$ (left) or $0.5^{\circ}$ (right) radius circle centered around the Galactic Center, for eight different combinations of parameters. We continue in this figure to assume that the entire Galactic Center Gamma-Ray Excess is generated by millisecond pulsars, but adopt either $\alpha = 1.5$ (top frames) or $\alpha = 2.0$ (bottom frames). In each frame, we show results for $E_{\rm cut}=30 \, {\rm TeV}$ or $100 \, {\rm TeV}$, and using our ``min'' or ``max'' model for the radiation and magnetic fields. As before, we have adopted $\eta =0.1$, $\langle \eta_{\rm GeV}\rangle =0.12$, $f_{\rm beam}=0.5$, and the spectra shown are as averaged over a $3^{\circ}$ (left) or $0.5^{\circ}$ (right) radius circle centered around the Galactic Center.}
\label{spectra2}
\end{figure*}

\begin{figure*}
\includegraphics[width=3.0in,angle=0]{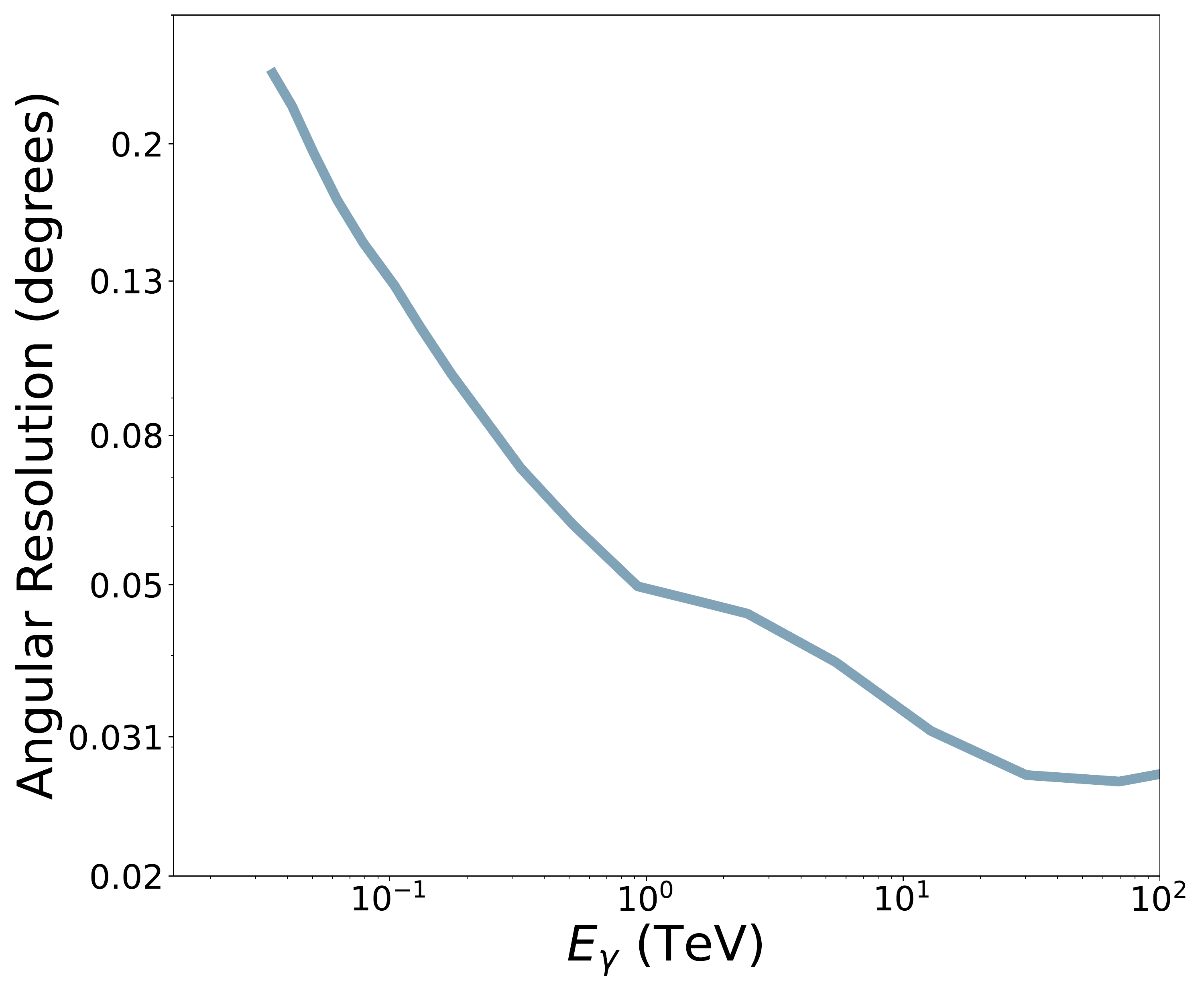} 
\includegraphics[width=3.0in,angle=0]{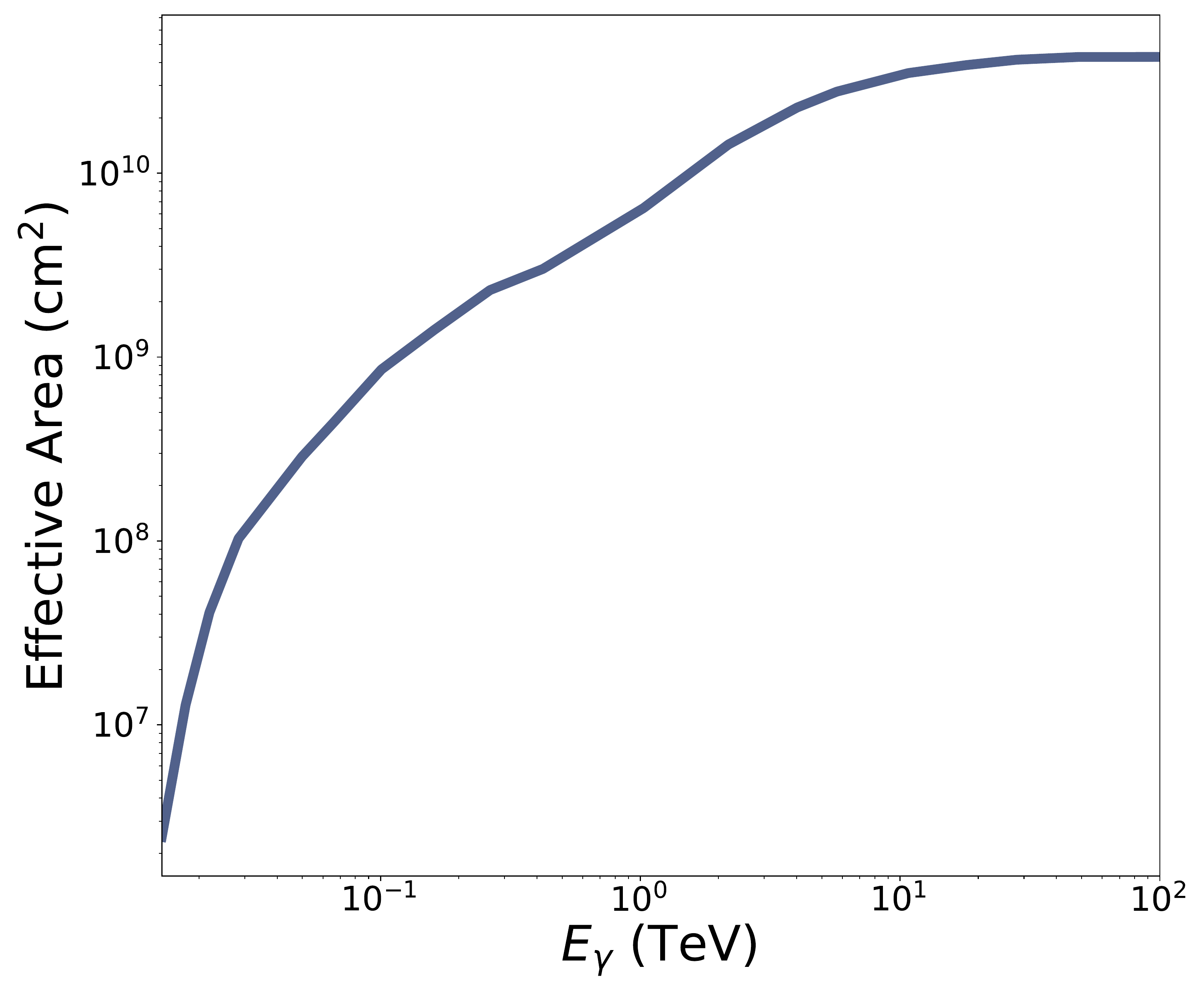}
\caption{The 68\% containment radius (left) and effective area (right) of CTA as a function of gamma-ray energy~\cite{bookCTA}.}
\label{astrogam}
\end{figure*}

\section{Observations With The Cherenkov Telescope Array}

The Cherenkov Telescope Array (CTA) is a next-generation ground-based gamma-ray telescope, designed to be sensitive to photons in the energy range of 20 GeV to 300 TeV~\cite{2019ICRC...36..733M}. Although CTA will occupy two sites, we will focus here on the southern hemisphere array, located in Paranal, Chile. This southern array will consist of 14 medium-sized and 37 small-sized telescopes, covering an area of approximately 4 km$^2$. This instrument is expected to have angular resolution on the order of two arcminutes, and energy resolution better than 10\%.

To evaluate the sensitivity of CTA to a population of TeV halos associated with millisecond pulsars in the Inner Galaxy, we have created a series of simulated data sets based on the proposed design of CTA, and analyzed this simulated data utilizing a set of spatial templates. Such template-based analyses are extremely powerful in that they allow us to simultaneously exploit both spectral and morphological distinctions between the signal being searched for, and the various astrophysical backgrounds that are present. In addition to the TeV halo template, our analysis includes spatial templates associated with the processes of pion production, bremsstrahlung, and (non-TeV halo) inverse Compton scattering, as well a template for the point sources described by the Fermi 4FGL-DR2 catalog~\cite{Fermi-LAT:2019yla}, and a template that is isotropic throughout our region-of-interest (a $3^{\circ}$ radius circle centered on the Galactic Center) which accounts for the emission associated with the Fermi Bubbles~\cite{Fermi-LAT:2014sfa} and the extragalactic gamma-ray background~\cite{Fermi-LAT:2014ryh}, as well as misidentified cosmic rays.\footnote{For the rate and spectrum of misidentified cosmic-ray, we use the background rate given at \url{https://www.cta-observatory.org/science/ctao-performance/} for CTA's Southern Array. Note that this background significantly exceeds those associated with the Fermi Bubbles and the extragalactic gamma-ray background.} The templates associated with pion production, bremsstrahlung, and inverse Compton scattering were each generated using the publicly available code, GALPROP~\cite{2011CoPhC.182.1156V,2005ApJ...622..759G}.\footnote{In using GALPROP, we have adopted the default parameters from GALPROP WebRun, \url{https://galprop.stanford.edu/webrun.php}, which have been selected to reproduce a wide variety of cosmic-ray and gamma-ray observations.} These templates are shown in Fig.~\ref{alltemplates}, and the gamma-ray spectrum associated with each of these components is shown in Fig.~\ref{spectra1}. In Fig.~\ref{spectra2}, we show the spectrum associated with the TeV halo template for a selection of different parameter choices. Note that in modeling the emission associated with the Fermi bubbles, point sources, and the extragalactic gamma-ray background, we are forced to extrapolate to energies well above those measured by Fermi, as shown in Fig.~\ref{spectra1}.

To produce a simulated data set, we first convolve each of the templates by the point spread function of CTA, which we treat as a gaussian with a 68\% containment radius as given in the left frame of Fig.~\ref{astrogam}~\cite{bookCTA}. Because of CTA's very high angular resolution, this convolution has very little effect on our results. We divide our region-of-interest (taken to be a $3^{\circ}$ radius circle around the Galactic Center) into 0.2098 square degree HEALPix bins, corresponding to $N_{\rm side}=128$. We also divide the spectrum into 10 energy bins per decade, spanning 10 GeV to 100 TeV. Then, after summing the templates described above, we calculate the mean number of events in a given angular and energy bin. This is done by multiplying the flux in that bin by the effective area of CTA (as given in the right frame of Fig.~\ref{astrogam})~\cite{bookCTA}, and by 20 hours of observation time. For each  bin, we then draw from a Poisson distribution using the mean number of events predicted by our model to obtain the simulated number of events in that bin. Using a simulated data set, we then calculate the likelihood as a function of our model's parameters. These parameters consist of the normalizations of each of our templates, in each energy bin. This procedure allows us to place constraints on the spectra of each of our background components, as well as on the spectrum of the emission from TeV halos associated with millisecond pulsars. In presenting our projected constraints, we show results as averaged over five sets of simulated data, to reduce the presence of statistical variation.\footnote{We use the publicly available MINUIT algorithm~\cite{iminuit} to perform our likelihood analysis. Given that MINUIT can occasionally identify false minima of the likelihood landscape, we also utilize the PyMultiNest package~\cite{Buchner:2014nha} to test the robustness of our results by searching for the global minimum in case it was not encountered in our MINUIT scan.}

\section{Projected Sensitivity}

\begin{figure}
\includegraphics[width=3.25in,angle=0]{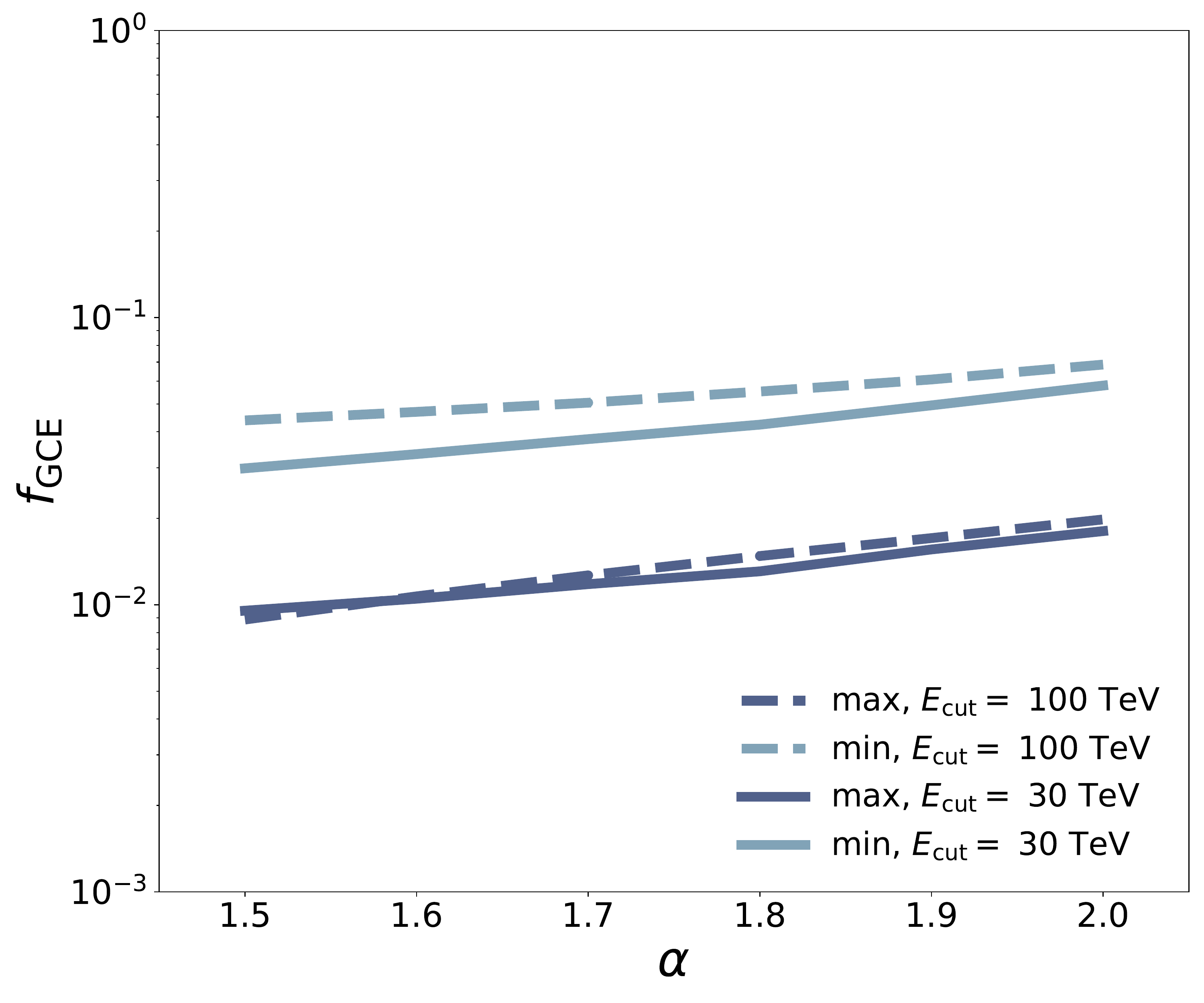}
\caption{The projected upper limit from CTA on the fraction of the Galactic Center Gamma-Ray Excess that is generated by millisecond pulsars, $f_{\rm GCE}$, under the assumption that no TeV halos are present in the Inner Galaxy. This result corresponds to 20 hours of observation of a $3^{\circ}$ radius circle around the Galactic Center, and we have adopted 
$\eta / (\langle \eta_{\rm GeV}\rangle f_{\mathrm{beam}}) = 1.67$. If no TeV halo emission is present in the Inner Galaxy, we expect CTA to be able to place extremely stringent constraints on the fraction of the excess that is generated by millisecond pulsars, at a level of approximately $f_{\rm GCE} \lsim (1-7)\%$.}
\label{constraintnopulsars}
\end{figure}

\begin{figure*}
\includegraphics[width=6.0in,angle=0]{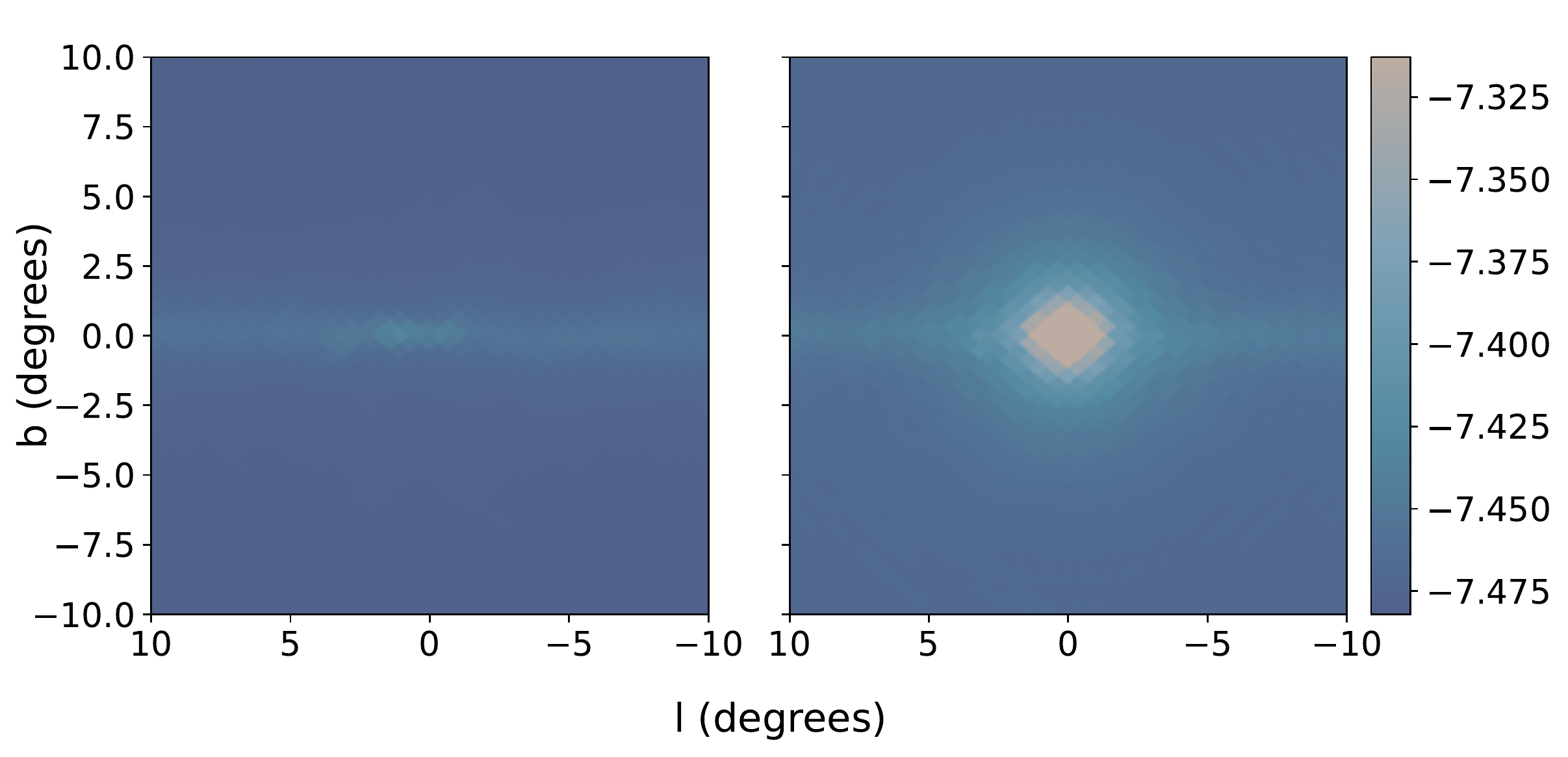}
\includegraphics[width=6.0in,angle=0]{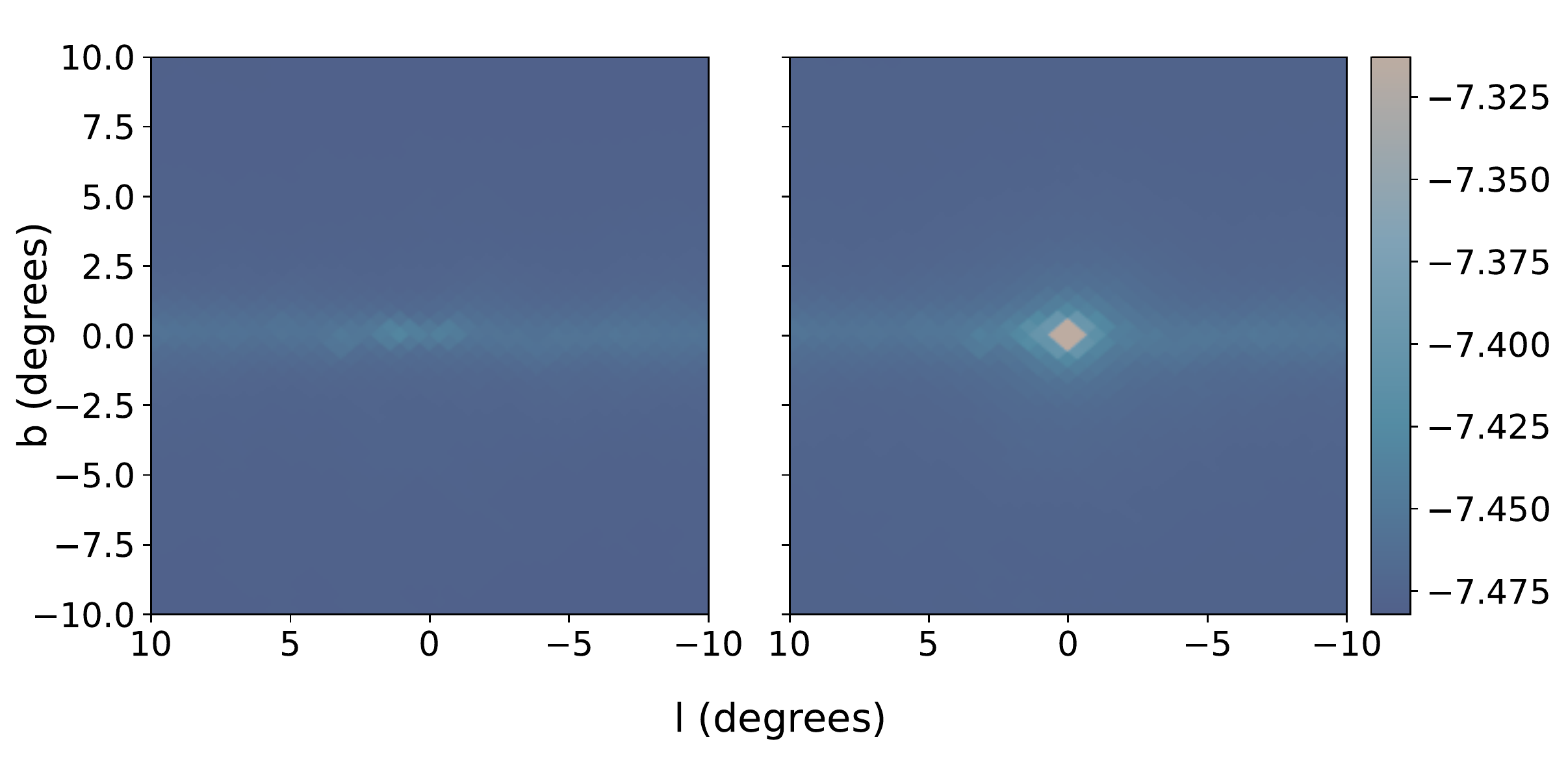}
\caption{The total gamma-ray emission predicted from our background model, including misidentified cosmic-rays (left), and from the background model plus the emission expected from TeV halos in the Inner Galaxy (right), assuming either that the entire Galactic Center Gamma-Ray Excess, or 20\% of it, is generated by millisecond pulsars, $f_{\rm GCE}=1$, (top) or $f_{\rm GCE}=0.2$ (bottom).  Here, we have adopted $\alpha = 1.5$, $E_{\rm cut}=30 \, {\rm TeV}$, $\eta =0.1$, $\langle \eta_{\rm GeV}\rangle =0.12$, $f_{\rm beam}=0.5$, and our ``min'' model for the radiation and magnetic fields. Each image represents the value of $\int dE_{\gamma} E_{\gamma } dN_{\gamma}/dE_{\gamma}$, integrated above 1 TeV, and in units of $\log_{10}$(TeV/cm$^2$/s/sr).}
\label{templatesum}
\end{figure*}

\begin{figure*}
\includegraphics[width=3.5in,angle=0]{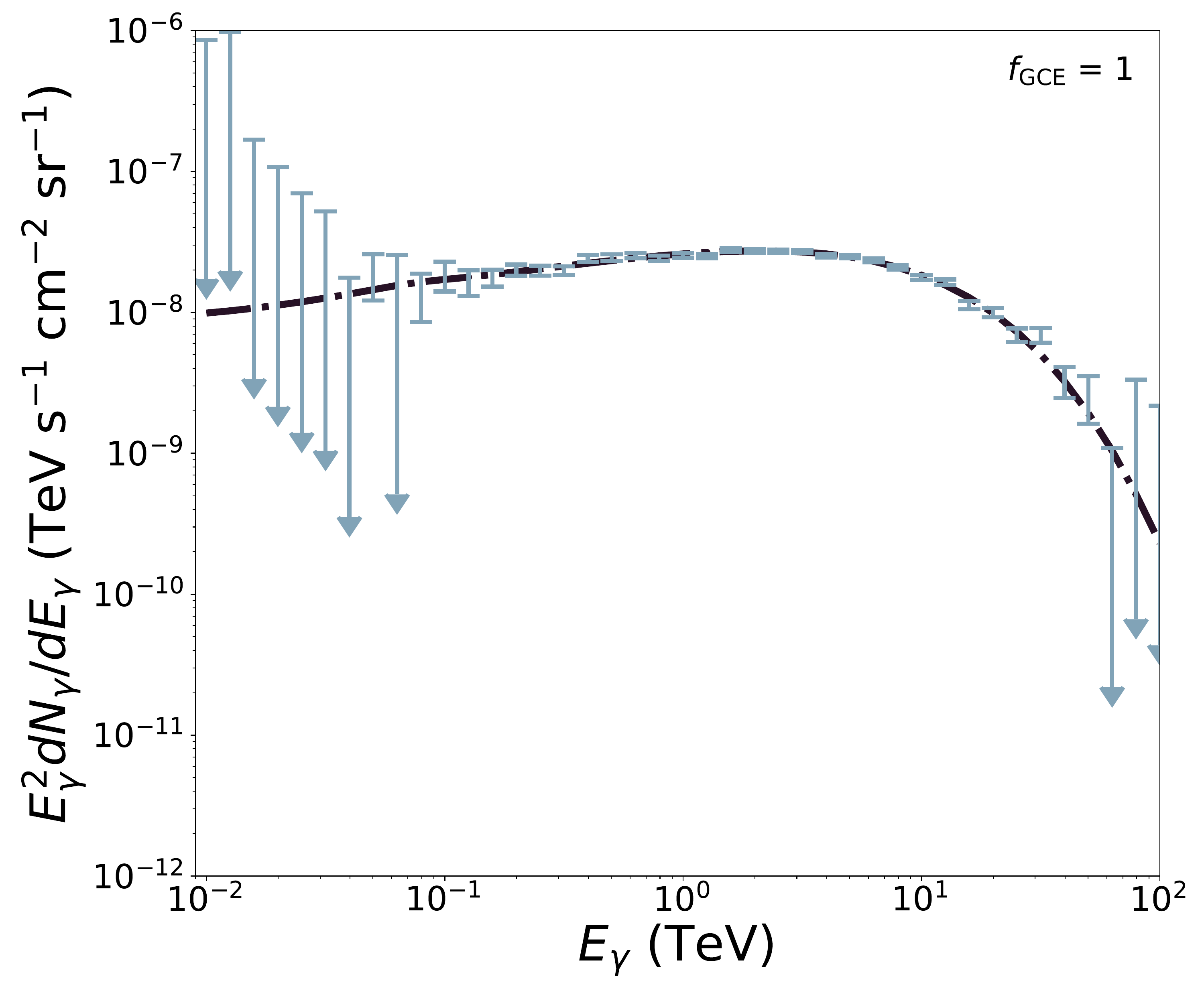}
\includegraphics[width=3.5in,angle=0]{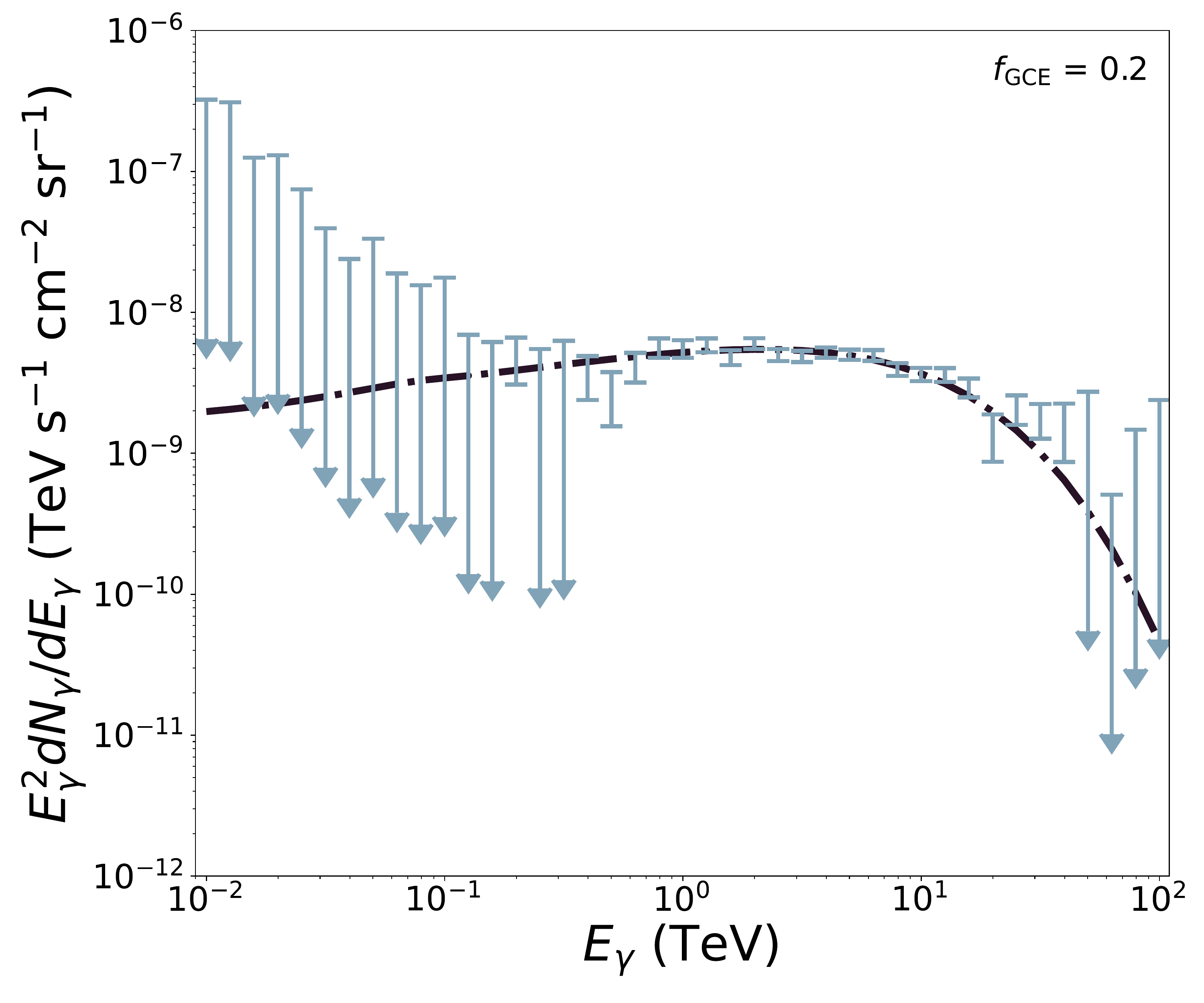}
\includegraphics[width=3.5in,angle=0]{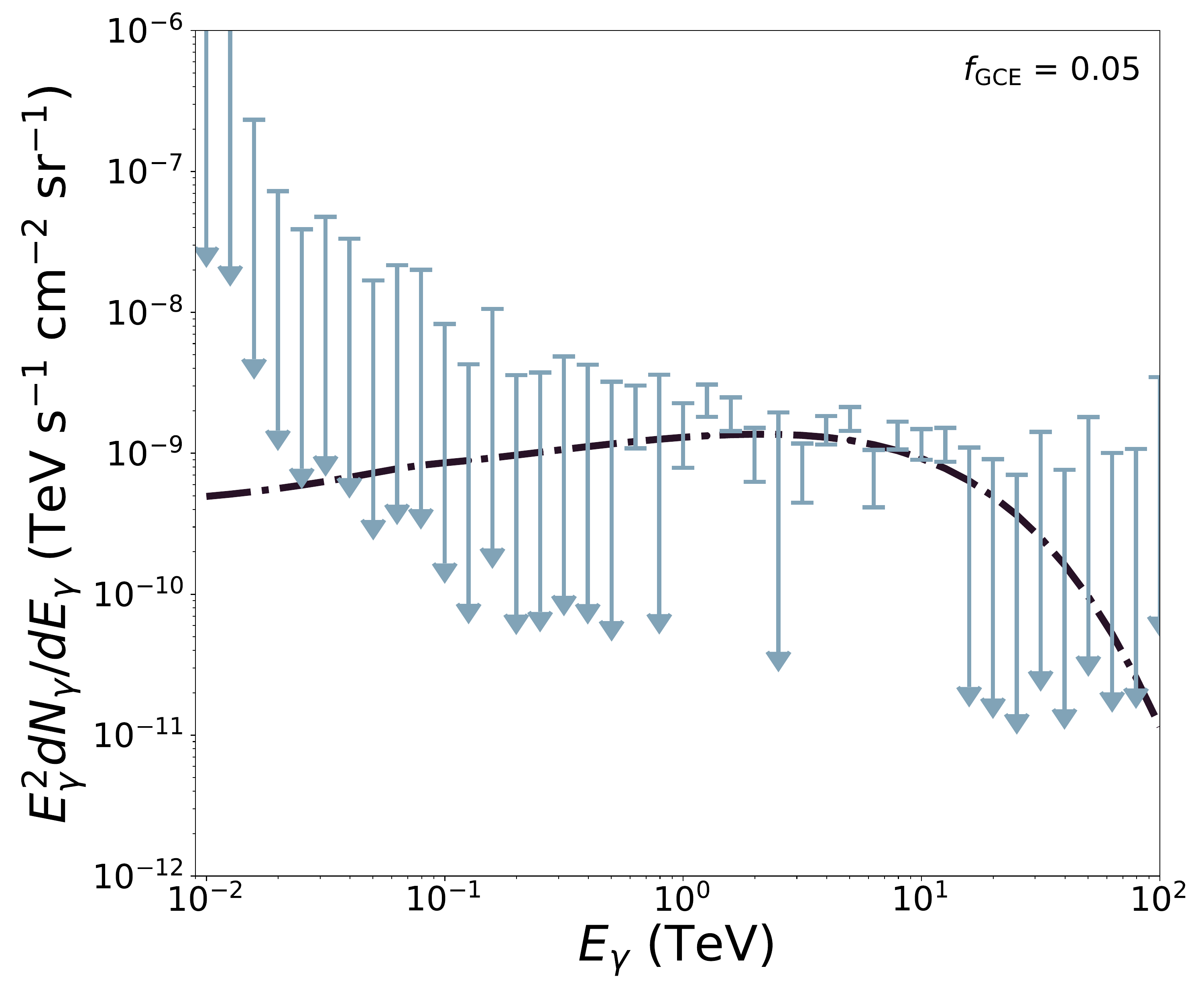}
\caption{The ability of CTA, after 20 hours of observation, to measure the spectrum of the gamma-ray emission from an Inner Galaxy population of TeV halos, for values of $f_{\rm GCE} = 1, 0.2$, or 0.05, and in each case adopting $\alpha = 1.5$, $E_{\rm cut}=30 \, {\rm TeV}$, $\eta =0.1$, $\langle \eta_{\rm GeV}\rangle =0.12$, $f_{\rm beam}=0.5$, and our ``min'' model for the radiation and magnetic fields. The gamma-ray emission from a centrally-located TeV halo population would be clearly identified and measured by CTA, even if only a relatively small fraction of the Galactic Center Gamma-Ray is generated by millisecond pulsars.}
\label{fig:spectraprojection}
\end{figure*}

We begin by generating a series of simulated data sets under the assumption that no TeV halos are present in the Inner Galaxy. We then perform our template analysis on this simulated data to place constraints on the TeV halo population. More specifically, we place an upper limit on the quantity, $\dot{E}_{\rm tot} \, \eta$, as a function of $\alpha$ and $E_{\rm cut}$. Equivalently, we can use Eq.~\ref{fGCE} to instead express this result as an upper limit on the fraction of the Galactic Center Gamma-Ray Excess that is generated by millisecond pulsars, $f_{\rm GCE}$, for a given choice of $\eta / (\langle \eta_{\rm GeV}\rangle f_{\mathrm{beam}})$.

The main result from this first analysis is shown in Fig.~\ref{constraintnopulsars}, where we plot the upper limit on $f_{\rm GCE}$, for the case of $\eta / (\langle \eta_{\rm GeV}\rangle f_{\mathrm{beam}}) = 0.1/(0.12 \times 0.5) = 1.67$. From this figure, we conclude that if no TeV halo emission is present in the Inner Galaxy, we can expect CTA to place an extremely stringent constraint on the fraction of the Galactic Center Gamma-Ray Excess that is generated by millisecond pulsars, at a level of approximately $f_{\rm GCE} \lsim (1-7)\%$. This projected limit is able to be so restrictive simply because the emission from the TeV halo population is incredibly bright in this case. This is illustrated in the spectra shown in Figs.~\ref{spectra1} and~\ref{spectra2}, as well as in Fig.~\ref{templatesum}, where we show a sky map of the gamma-ray intensity in the Inner Galaxy above 0.1 TeV for the case of our background model, and for our background model plus the TeV halo emission, normalized to $f_{\rm GCE}=1$ or 0.2 for $\eta / (\langle \eta_{\rm GeV}\rangle f_{\mathrm{beam}}) = 1.67$. The presence of TeV halos in the Inner Galaxy dramatically changes the morphology and spectrum of the very high-energy gamma-ray emission from this region, and should be easily detectable after even a brief observation by CTA.

For our second analysis, we have generated a series of simulated data sets assuming that TeV halos {\it are} present in the Inner Galaxy. The ability of CTA to measure the spectrum of the gamma-ray emission from these Inner Galaxy TeV halos is shown in Fig.~\ref{fig:spectraprojection}, for values of $f_{\rm GCE} = 1, 0.2$, or 0.05 (and in each case adopting $\alpha = 1.5$, $E_{\rm cut}=30 \, {\rm TeV}$, $\eta =0.1$, $\langle \eta_{\rm GeV}\rangle =0.12$, $f_{\rm beam}=0.5$, and our ``min'' model for the radiation and magnetic fields). Note that these figures depict the results of an individual simulated data set, explaining the bin-to-bin variations seen in the projected error bars. These results demonstrate that CTA should be able to clearly identify and measure the gamma-ray emission from a centrally-located TeV halo population, even if only a relatively small fraction of the Galactic Center Gamma-Ray is generated by millisecond pulsars.

\section{Discussion}

Measurements of the very high-energy gamma-ray emission from the Inner Galaxy have already been reported by the HESS Collaboration ~\cite{HESS:2016pst}. That study, however, produced spectra only from the innermost $0.5^{\circ}$ around the Galactic Center. In that region, the observed emission is approximately equal to that expected from TeV halos in a scenario in which the entire Galactic Center Gamma-Ray Excess originates from millisecond pulsars~\cite{Hooper:2018fih} (see also, Ref.~\cite{Hooper:2017rzt}). This information allows us to place an upper limit of roughly $f_{\rm GCE} \, \eta / (\langle \eta_{\rm GeV}\rangle f_{\mathrm{beam}}) \lsim  \mathcal{O}(1-3)$ on our TeV halo/MSP parameters. Furthermore, it has been argued by the HESS Collaboration that this emission is correlated with the observed distribution of molecular gas, suggesting a hadronic origin~\cite{HESS:2016pst}, and thereby further limiting how much of this emission could originate from  TeV halos. With CTA's greater sensitivity and larger field-of-view, it should be possible to rigorously test this interpretation, in particular by measuring the spectrum and intensity of the gamma-ray emission from the regions several degrees north or south of the Galactic Center.

A number of assumptions have gone into our analysis which could potentially impact our conclusions. First, we have assumed that the emission from TeV halos has the same angular distribution as that from the GeV-scale emission from MSPs (and thus traces the morphology of the Galactic Center Gamma-Ray Excess). The transport of electrons and positrons could, however, broaden the TeV halo signal, at least in principle. The TeV halos observed by HAWC are extended on scales of roughly $\sim$20-30~pc, corresponding to an angular extent of $\sim 0.14^{\circ}-0.21^{\circ}$ for sources located near the Galactic Center. This has motivated our choice for the size of the angular bins used in our study, 0.2098 square degrees, which is slightly larger than the size of a typical TeV halo, limiting the impact of particle transport on our signal. At this time, however, it has not yet been established whether diffusion is inhibited within the volumes surrounding MSPs, as it is observed to be around young and middle-aged pulsars~\cite{Evoli:2018aza,Fang:2019iym}. If diffusion is more efficient around MSPs, the signal of these TeV halos could be broadened by up to $\sim \mathcal{O}(1^{\circ})$, while still leaving the total flux of gamma-ray emission from these sources approximately unchanged. Future measurements of the TeV halos surrounding nearby MSPs will inform the shape of our TeV halo template.

Second, if the magnetic fields are very large in the Inner Galaxy, most of the energy injected into very high-energy electron-positron pairs could emitted as synchrotron emission, rather than as inverse Compton, suppressing the signal at TeV-scale energies. While this is unlikely to be an important factor over most of our region-of-interest, it could be significant in the innermost volume around the Galactic Center, where magnetic fields could potentially be quite large. This scenario, however, is limited by radio observations of the Inner Galaxy~\cite{Crocker:2010xc}. If millisecond pulsars generate all or most of the Galactic Center Gamma-Ray Excess, their TeV halos will produce electrons and positrons which lose most of the energy to inverse Compton scattering~\cite{Hooper:2018fih}.

Third, throughout this study, we have scaled our results to $\eta / (\langle \eta_{\rm GeV}\rangle f_{\mathrm{beam}}) = 0.1/(0.12\times 0.5)=1.67$. As the predicted intensity of the TeV halo emission scales with this combination of parameters, our constraints would be less stringent if the true value of $\eta / (\langle \eta_{\rm GeV}\rangle f_{\mathrm{beam}})$ were significantly smaller than our default value. Fortunately, CTA should be able to measure the value (or distribution) of $\eta$ across a large number of pulsars, including many with millisecond-scale periods, reducing this uncertainty considerably. In light of this, we expect that the uncertainty on $\langle \eta_{\rm GeV}\rangle f_{\mathrm{beam}}$ will not qualitatively impact our conclusions.

Lastly, while the conclusions reached in this study are in broad agreement with those presented in Ref.~\cite{Macias:2021boz}, that paper's projections for CTA's sensitivity to a centrally-located TeV halo population are somewhat more modest than those reported here. While the precise origin of this difference is not entirely clear, we note that the authors of Ref.~\cite{Macias:2021boz} have adopted a model for the gamma-ray emission associated with the Fermi Bubbles that significantly increases in brightness at low galactic latitudes. In contrast, we have used the fact that the brightness of the Bubbles is approximately uniform between $10^{\circ} < |b| < 50^{\circ}$~\cite{Fermi-LAT:2014sfa} to motivate the assumption that this emission is comparably bright at $|b| < 10^{\circ}$. Regardless of this distinction, both groups reach the same conclusion that CTA should be able to clearly identify the emission associated from TeV halos if the Galactic Center Gamma-Ray Excess originates from millisecond pulsars.

\section{Summary and Conclusions}
 
The Galactic Center Gamma-Ray Excess, as observed by Fermi, has a spectrum, angular distribution, and overall intensity that is in good agreement with that expected to be produced by annihilating dark matter particles. The leading alternative to this interpretation is that this signal is instead generated by a large number of unresolved millisecond pulsars. If millisecond pulsars are responsible for this signal, however, they should also produce detectable levels of TeV-scale emission, as recent observations indicate that pulsars appear to be universally surrounded by bright, spatially-extended, multi-TeV emitting regions known as ``TeV halos''.

In this study, we have performed a template-based analysis of simulated data to assess the ability of the Cherenkov Telescope Array (CTA) to identify and measure the very high-energy gamma-ray emission from a population of millisecond pulsars surrounding the Galactic Center. When the intensity of this emission is normalized to the measurements of other pulsars by HAWC and LHAASO, we find that if the Galactic Center Gamma-Ray Excess originates from pulsars, this source population should produce a very bright flux of gamma rays, dominating the total TeV-scale emission from the direction of the Inner Galaxy. Such a signal would be easily detectable by CTA, even after a relatively brief period of observation. If such a pulsar population is present in the Inner Galaxy, we conclude that CTA should be able to clearly identify the TeV halo emission associated with these sources. If this emission is not observed by CTA, this would strongly rule out the hypothesis that the Galactic Center Gamma-Ray Excess originates from millisecond pulsars.

\begin{acknowledgments}  

CK is supported by the National Science Foundation Graduate Research Fellowship Program under grant No.~DGE-1746045, and by the U.S. Department of Energy Office of Science Graduate Student Research (SCGSR) program under contract number DE‐SC0014664. DH is supported by the Fermi Research Alliance, LLC under Contract No.~DE-AC02-07CH11359 with the U.S. Department of Energy, Office of Science, Office of High Energy Physics. TL is partially supported by the Swedish Research Council under contract 2019-05135, the Swedish National Space Agency under contract 117/19 and the European Research Council under grant 742104. CK would like to thank Elyssa Cotter for helpful conversations.

\end{acknowledgments}

\bibliography{pulsarspref}

\begin{thebibliography}{81}%
\makeatletter
\providecommand \@ifxundefined [1]{%
 \@ifx{#1\undefined}
}%
\providecommand \@ifnum [1]{%
 \ifnum #1\expandafter \@firstoftwo
 \else \expandafter \@secondoftwo
 \fi
}%
\providecommand \@ifx [1]{%
 \ifx #1\expandafter \@firstoftwo
 \else \expandafter \@secondoftwo
 \fi
}%
\providecommand \natexlab [1]{#1}%
\providecommand \enquote  [1]{``#1''}%
\providecommand \bibnamefont  [1]{#1}%
\providecommand \bibfnamefont [1]{#1}%
\providecommand \citenamefont [1]{#1}%
\providecommand \href@noop [0]{\@secondoftwo}%
\providecommand \href [0]{\begingroup \@sanitize@url \@href}%
\providecommand \@href[1]{\@@startlink{#1}\@@href}%
\providecommand \@@href[1]{\endgroup#1\@@endlink}%
\providecommand \@sanitize@url [0]{\catcode `\\12\catcode `\$12\catcode
  `\&12\catcode `\#12\catcode `\^12\catcode `\_12\catcode `\%12\relax}%
\providecommand \@@startlink[1]{}%
\providecommand \@@endlink[0]{}%
\providecommand \url  [0]{\begingroup\@sanitize@url \@url }%
\providecommand \@url [1]{\endgroup\@href {#1}{\urlprefix }}%
\providecommand \urlprefix  [0]{URL }%
\providecommand \Eprint [0]{\href }%
\providecommand \doibase [0]{http://dx.doi.org/}%
\providecommand \selectlanguage [0]{\@gobble}%
\providecommand \bibinfo  [0]{\@secondoftwo}%
\providecommand \bibfield  [0]{\@secondoftwo}%
\providecommand \translation [1]{[#1]}%
\providecommand \BibitemOpen [0]{}%
\providecommand \bibitemStop [0]{}%
\providecommand \bibitemNoStop [0]{.\EOS\space}%
\providecommand \EOS [0]{\spacefactor3000\relax}%
\providecommand \BibitemShut  [1]{\csname bibitem#1\endcsname}%
\let\auto@bib@innerbib\@empty
%</preamble>
\bibitem [{\citenamefont {Goodenough}\ and\ \citenamefont
  {Hooper}(2009)}]{Goodenough:2009gk}%
  \BibitemOpen
  \bibfield  {author} {\bibinfo {author} {\bibfnamefont {L.}~\bibnamefont
  {Goodenough}}\ and\ \bibinfo {author} {\bibfnamefont {D.}~\bibnamefont
  {Hooper}},\ }\href@noop {} {\  (\bibinfo {year} {2009})},\ \Eprint
  {http://arxiv.org/abs/0910.2998} {arXiv:0910.2998 [hep-ph]} \BibitemShut
  {NoStop}%
\bibitem [{\citenamefont {Hooper}\ and\ \citenamefont
  {Goodenough}(2011)}]{Hooper:2010mq}%
  \BibitemOpen
  \bibfield  {author} {\bibinfo {author} {\bibfnamefont {D.}~\bibnamefont
  {Hooper}}\ and\ \bibinfo {author} {\bibfnamefont {L.}~\bibnamefont
  {Goodenough}},\ }\href {\doibase 10.1016/j.physletb.2011.02.029} {\bibfield
  {journal} {\bibinfo  {journal} {Phys. Lett. B}\ }\textbf {\bibinfo {volume}
  {697}},\ \bibinfo {pages} {412} (\bibinfo {year} {2011})},\ \Eprint
  {http://arxiv.org/abs/1010.2752} {arXiv:1010.2752 [hep-ph]} \BibitemShut
  {NoStop}%
\bibitem [{\citenamefont {Hooper}\ and\ \citenamefont
  {Linden}(2011)}]{Hooper:2011ti}%
  \BibitemOpen
  \bibfield  {author} {\bibinfo {author} {\bibfnamefont {D.}~\bibnamefont
  {Hooper}}\ and\ \bibinfo {author} {\bibfnamefont {T.}~\bibnamefont
  {Linden}},\ }\href {\doibase 10.1103/PhysRevD.84.123005} {\bibfield
  {journal} {\bibinfo  {journal} {Phys. Rev. D}\ }\textbf {\bibinfo {volume}
  {84}},\ \bibinfo {pages} {123005} (\bibinfo {year} {2011})},\ \Eprint
  {http://arxiv.org/abs/1110.0006} {arXiv:1110.0006 [astro-ph.HE]} \BibitemShut
  {NoStop}%
\bibitem [{\citenamefont {Daylan}\ \emph {et~al.}(2016)\citenamefont {Daylan},
  \citenamefont {Finkbeiner}, \citenamefont {Hooper}, \citenamefont {Linden},
  \citenamefont {Portillo}, \citenamefont {Rodd},\ and\ \citenamefont
  {Slatyer}}]{Daylan:2014rsa}%
  \BibitemOpen
  \bibfield  {author} {\bibinfo {author} {\bibfnamefont {T.}~\bibnamefont
  {Daylan}}, \bibinfo {author} {\bibfnamefont {D.~P.}\ \bibnamefont
  {Finkbeiner}}, \bibinfo {author} {\bibfnamefont {D.}~\bibnamefont {Hooper}},
  \bibinfo {author} {\bibfnamefont {T.}~\bibnamefont {Linden}}, \bibinfo
  {author} {\bibfnamefont {S.~K.~N.}\ \bibnamefont {Portillo}}, \bibinfo
  {author} {\bibfnamefont {N.~L.}\ \bibnamefont {Rodd}}, \ and\ \bibinfo
  {author} {\bibfnamefont {T.~R.}\ \bibnamefont {Slatyer}},\ }\href {\doibase
  10.1016/j.dark.2015.12.005} {\bibfield  {journal} {\bibinfo  {journal} {Phys.
  Dark Univ.}\ }\textbf {\bibinfo {volume} {12}},\ \bibinfo {pages} {1}
  (\bibinfo {year} {2016})},\ \Eprint {http://arxiv.org/abs/1402.6703}
  {arXiv:1402.6703 [astro-ph.HE]} \BibitemShut {NoStop}%
\bibitem [{\citenamefont {Calore}\ \emph
  {et~al.}(2015{\natexlab{a}})\citenamefont {Calore}, \citenamefont {Cholis},\
  and\ \citenamefont {Weniger}}]{Calore:2014xka}%
  \BibitemOpen
  \bibfield  {author} {\bibinfo {author} {\bibfnamefont {F.}~\bibnamefont
  {Calore}}, \bibinfo {author} {\bibfnamefont {I.}~\bibnamefont {Cholis}}, \
  and\ \bibinfo {author} {\bibfnamefont {C.}~\bibnamefont {Weniger}},\ }\href
  {\doibase 10.1088/1475-7516/2015/03/038} {\bibfield  {journal} {\bibinfo
  {journal} {JCAP}\ }\textbf {\bibinfo {volume} {03}},\ \bibinfo {pages} {038}
  (\bibinfo {year} {2015}{\natexlab{a}})},\ \Eprint
  {http://arxiv.org/abs/1409.0042} {arXiv:1409.0042 [astro-ph.CO]} \BibitemShut
  {NoStop}%
\bibitem [{\citenamefont {Calore}\ \emph
  {et~al.}(2015{\natexlab{b}})\citenamefont {Calore}, \citenamefont {Cholis},
  \citenamefont {McCabe},\ and\ \citenamefont {Weniger}}]{Calore:2014nla}%
  \BibitemOpen
  \bibfield  {author} {\bibinfo {author} {\bibfnamefont {F.}~\bibnamefont
  {Calore}}, \bibinfo {author} {\bibfnamefont {I.}~\bibnamefont {Cholis}},
  \bibinfo {author} {\bibfnamefont {C.}~\bibnamefont {McCabe}}, \ and\ \bibinfo
  {author} {\bibfnamefont {C.}~\bibnamefont {Weniger}},\ }\href {\doibase
  10.1103/PhysRevD.91.063003} {\bibfield  {journal} {\bibinfo  {journal} {Phys.
  Rev. D}\ }\textbf {\bibinfo {volume} {91}},\ \bibinfo {pages} {063003}
  (\bibinfo {year} {2015}{\natexlab{b}})},\ \Eprint
  {http://arxiv.org/abs/1411.4647} {arXiv:1411.4647 [hep-ph]} \BibitemShut
  {NoStop}%
\bibitem [{\citenamefont {Ackermann}\ \emph {et~al.}(2017)\citenamefont
  {Ackermann} \emph {et~al.}}]{Fermi-LAT:2017opo}%
  \BibitemOpen
  \bibfield  {author} {\bibinfo {author} {\bibfnamefont {M.}~\bibnamefont
  {Ackermann}} \emph {et~al.} (\bibinfo {collaboration} {Fermi-LAT}),\ }\href
  {\doibase 10.3847/1538-4357/aa6cab} {\bibfield  {journal} {\bibinfo
  {journal} {Astrophys. J.}\ }\textbf {\bibinfo {volume} {840}},\ \bibinfo
  {pages} {43} (\bibinfo {year} {2017})},\ \Eprint
  {http://arxiv.org/abs/1704.03910} {arXiv:1704.03910 [astro-ph.HE]}
  \BibitemShut {NoStop}%
\bibitem [{\citenamefont {Di~Mauro}(2021)}]{DiMauro:2021raz}%
  \BibitemOpen
  \bibfield  {author} {\bibinfo {author} {\bibfnamefont {M.}~\bibnamefont
  {Di~Mauro}},\ }\href {\doibase 10.1103/PhysRevD.103.063029} {\bibfield
  {journal} {\bibinfo  {journal} {Phys. Rev. D}\ }\textbf {\bibinfo {volume}
  {103}},\ \bibinfo {pages} {063029} (\bibinfo {year} {2021})},\ \Eprint
  {http://arxiv.org/abs/2101.04694} {arXiv:2101.04694 [astro-ph.HE]}
  \BibitemShut {NoStop}%
\bibitem [{\citenamefont {Cholis}\ \emph {et~al.}(2022)\citenamefont {Cholis},
  \citenamefont {Zhong}, \citenamefont {McDermott},\ and\ \citenamefont
  {Surdutovich}}]{Cholis:2021rpp}%
  \BibitemOpen
  \bibfield  {author} {\bibinfo {author} {\bibfnamefont {I.}~\bibnamefont
  {Cholis}}, \bibinfo {author} {\bibfnamefont {Y.-M.}\ \bibnamefont {Zhong}},
  \bibinfo {author} {\bibfnamefont {S.~D.}\ \bibnamefont {McDermott}}, \ and\
  \bibinfo {author} {\bibfnamefont {J.~P.}\ \bibnamefont {Surdutovich}},\
  }\href {\doibase 10.1103/PhysRevD.105.103023} {\bibfield  {journal} {\bibinfo
   {journal} {Phys. Rev. D}\ }\textbf {\bibinfo {volume} {105}},\ \bibinfo
  {pages} {103023} (\bibinfo {year} {2022})},\ \Eprint
  {http://arxiv.org/abs/2112.09706} {arXiv:2112.09706 [astro-ph.HE]}
  \BibitemShut {NoStop}%
\bibitem [{\citenamefont {Berlin}\ \emph
  {et~al.}(2014{\natexlab{a}})\citenamefont {Berlin}, \citenamefont {Hooper},\
  and\ \citenamefont {McDermott}}]{Berlin:2014tja}%
  \BibitemOpen
  \bibfield  {author} {\bibinfo {author} {\bibfnamefont {A.}~\bibnamefont
  {Berlin}}, \bibinfo {author} {\bibfnamefont {D.}~\bibnamefont {Hooper}}, \
  and\ \bibinfo {author} {\bibfnamefont {S.~D.}\ \bibnamefont {McDermott}},\
  }\href {\doibase 10.1103/PhysRevD.89.115022} {\bibfield  {journal} {\bibinfo
  {journal} {Phys. Rev. D}\ }\textbf {\bibinfo {volume} {89}},\ \bibinfo
  {pages} {115022} (\bibinfo {year} {2014}{\natexlab{a}})},\ \Eprint
  {http://arxiv.org/abs/1404.0022} {arXiv:1404.0022 [hep-ph]} \BibitemShut
  {NoStop}%
\bibitem [{\citenamefont {Izaguirre}\ \emph {et~al.}(2014)\citenamefont
  {Izaguirre}, \citenamefont {Krnjaic},\ and\ \citenamefont
  {Shuve}}]{Izaguirre:2014vva}%
  \BibitemOpen
  \bibfield  {author} {\bibinfo {author} {\bibfnamefont {E.}~\bibnamefont
  {Izaguirre}}, \bibinfo {author} {\bibfnamefont {G.}~\bibnamefont {Krnjaic}},
  \ and\ \bibinfo {author} {\bibfnamefont {B.}~\bibnamefont {Shuve}},\ }\href
  {\doibase 10.1103/PhysRevD.90.055002} {\bibfield  {journal} {\bibinfo
  {journal} {Phys. Rev. D}\ }\textbf {\bibinfo {volume} {90}},\ \bibinfo
  {pages} {055002} (\bibinfo {year} {2014})},\ \Eprint
  {http://arxiv.org/abs/1404.2018} {arXiv:1404.2018 [hep-ph]} \BibitemShut
  {NoStop}%
\bibitem [{\citenamefont {Ipek}\ \emph {et~al.}(2014)\citenamefont {Ipek},
  \citenamefont {McKeen},\ and\ \citenamefont {Nelson}}]{Ipek:2014gua}%
  \BibitemOpen
  \bibfield  {author} {\bibinfo {author} {\bibfnamefont {S.}~\bibnamefont
  {Ipek}}, \bibinfo {author} {\bibfnamefont {D.}~\bibnamefont {McKeen}}, \ and\
  \bibinfo {author} {\bibfnamefont {A.~E.}\ \bibnamefont {Nelson}},\ }\href
  {\doibase 10.1103/PhysRevD.90.055021} {\bibfield  {journal} {\bibinfo
  {journal} {Phys. Rev. D}\ }\textbf {\bibinfo {volume} {90}},\ \bibinfo
  {pages} {055021} (\bibinfo {year} {2014})},\ \Eprint
  {http://arxiv.org/abs/1404.3716} {arXiv:1404.3716 [hep-ph]} \BibitemShut
  {NoStop}%
\bibitem [{\citenamefont {Agrawal}\ \emph {et~al.}(2014)\citenamefont
  {Agrawal}, \citenamefont {Batell}, \citenamefont {Hooper},\ and\
  \citenamefont {Lin}}]{Agrawal:2014una}%
  \BibitemOpen
  \bibfield  {author} {\bibinfo {author} {\bibfnamefont {P.}~\bibnamefont
  {Agrawal}}, \bibinfo {author} {\bibfnamefont {B.}~\bibnamefont {Batell}},
  \bibinfo {author} {\bibfnamefont {D.}~\bibnamefont {Hooper}}, \ and\ \bibinfo
  {author} {\bibfnamefont {T.}~\bibnamefont {Lin}},\ }\href {\doibase
  10.1103/PhysRevD.90.063512} {\bibfield  {journal} {\bibinfo  {journal} {Phys.
  Rev. D}\ }\textbf {\bibinfo {volume} {90}},\ \bibinfo {pages} {063512}
  (\bibinfo {year} {2014})},\ \Eprint {http://arxiv.org/abs/1404.1373}
  {arXiv:1404.1373 [hep-ph]} \BibitemShut {NoStop}%
\bibitem [{\citenamefont {Berlin}\ \emph
  {et~al.}(2014{\natexlab{b}})\citenamefont {Berlin}, \citenamefont {Gratia},
  \citenamefont {Hooper},\ and\ \citenamefont {McDermott}}]{Berlin:2014pya}%
  \BibitemOpen
  \bibfield  {author} {\bibinfo {author} {\bibfnamefont {A.}~\bibnamefont
  {Berlin}}, \bibinfo {author} {\bibfnamefont {P.}~\bibnamefont {Gratia}},
  \bibinfo {author} {\bibfnamefont {D.}~\bibnamefont {Hooper}}, \ and\ \bibinfo
  {author} {\bibfnamefont {S.~D.}\ \bibnamefont {McDermott}},\ }\href {\doibase
  10.1103/PhysRevD.90.015032} {\bibfield  {journal} {\bibinfo  {journal} {Phys.
  Rev. D}\ }\textbf {\bibinfo {volume} {90}},\ \bibinfo {pages} {015032}
  (\bibinfo {year} {2014}{\natexlab{b}})},\ \Eprint
  {http://arxiv.org/abs/1405.5204} {arXiv:1405.5204 [hep-ph]} \BibitemShut
  {NoStop}%
\bibitem [{\citenamefont {Abdullah}\ \emph {et~al.}(2014)\citenamefont
  {Abdullah}, \citenamefont {DiFranzo}, \citenamefont {Rajaraman},
  \citenamefont {Tait}, \citenamefont {Tanedo},\ and\ \citenamefont
  {Wijangco}}]{Abdullah:2014lla}%
  \BibitemOpen
  \bibfield  {author} {\bibinfo {author} {\bibfnamefont {M.}~\bibnamefont
  {Abdullah}}, \bibinfo {author} {\bibfnamefont {A.}~\bibnamefont {DiFranzo}},
  \bibinfo {author} {\bibfnamefont {A.}~\bibnamefont {Rajaraman}}, \bibinfo
  {author} {\bibfnamefont {T.~M.~P.}\ \bibnamefont {Tait}}, \bibinfo {author}
  {\bibfnamefont {P.}~\bibnamefont {Tanedo}}, \ and\ \bibinfo {author}
  {\bibfnamefont {A.~M.}\ \bibnamefont {Wijangco}},\ }\href {\doibase
  10.1103/PhysRevD.90.035004} {\bibfield  {journal} {\bibinfo  {journal} {Phys.
  Rev. D}\ }\textbf {\bibinfo {volume} {90}},\ \bibinfo {pages} {035004}
  (\bibinfo {year} {2014})},\ \Eprint {http://arxiv.org/abs/1404.6528}
  {arXiv:1404.6528 [hep-ph]} \BibitemShut {NoStop}%
\bibitem [{\citenamefont {Martin}\ \emph {et~al.}(2014)\citenamefont {Martin},
  \citenamefont {Shelton},\ and\ \citenamefont {Unwin}}]{Martin:2014sxa}%
  \BibitemOpen
  \bibfield  {author} {\bibinfo {author} {\bibfnamefont {A.}~\bibnamefont
  {Martin}}, \bibinfo {author} {\bibfnamefont {J.}~\bibnamefont {Shelton}}, \
  and\ \bibinfo {author} {\bibfnamefont {J.}~\bibnamefont {Unwin}},\ }\href
  {\doibase 10.1103/PhysRevD.90.103513} {\bibfield  {journal} {\bibinfo
  {journal} {Phys. Rev. D}\ }\textbf {\bibinfo {volume} {90}},\ \bibinfo
  {pages} {103513} (\bibinfo {year} {2014})},\ \Eprint
  {http://arxiv.org/abs/1405.0272} {arXiv:1405.0272 [hep-ph]} \BibitemShut
  {NoStop}%
\bibitem [{\citenamefont {Alves}\ \emph {et~al.}(2014)\citenamefont {Alves},
  \citenamefont {Profumo}, \citenamefont {Queiroz},\ and\ \citenamefont
  {Shepherd}}]{Alves:2014yha}%
  \BibitemOpen
  \bibfield  {author} {\bibinfo {author} {\bibfnamefont {A.}~\bibnamefont
  {Alves}}, \bibinfo {author} {\bibfnamefont {S.}~\bibnamefont {Profumo}},
  \bibinfo {author} {\bibfnamefont {F.~S.}\ \bibnamefont {Queiroz}}, \ and\
  \bibinfo {author} {\bibfnamefont {W.}~\bibnamefont {Shepherd}},\ }\href
  {\doibase 10.1103/PhysRevD.90.115003} {\bibfield  {journal} {\bibinfo
  {journal} {Phys. Rev. D}\ }\textbf {\bibinfo {volume} {90}},\ \bibinfo
  {pages} {115003} (\bibinfo {year} {2014})},\ \Eprint
  {http://arxiv.org/abs/1403.5027} {arXiv:1403.5027 [hep-ph]} \BibitemShut
  {NoStop}%
\bibitem [{\citenamefont {Boehm}\ \emph {et~al.}(2014)\citenamefont {Boehm},
  \citenamefont {Dolan}, \citenamefont {McCabe}, \citenamefont {Spannowsky},\
  and\ \citenamefont {Wallace}}]{Boehm:2014hva}%
  \BibitemOpen
  \bibfield  {author} {\bibinfo {author} {\bibfnamefont {C.}~\bibnamefont
  {Boehm}}, \bibinfo {author} {\bibfnamefont {M.~J.}\ \bibnamefont {Dolan}},
  \bibinfo {author} {\bibfnamefont {C.}~\bibnamefont {McCabe}}, \bibinfo
  {author} {\bibfnamefont {M.}~\bibnamefont {Spannowsky}}, \ and\ \bibinfo
  {author} {\bibfnamefont {C.~J.}\ \bibnamefont {Wallace}},\ }\href {\doibase
  10.1088/1475-7516/2014/05/009} {\bibfield  {journal} {\bibinfo  {journal}
  {JCAP}\ }\textbf {\bibinfo {volume} {05}},\ \bibinfo {pages} {009} (\bibinfo
  {year} {2014})},\ \Eprint {http://arxiv.org/abs/1401.6458} {arXiv:1401.6458
  [hep-ph]} \BibitemShut {NoStop}%
\bibitem [{\citenamefont {Berlin}\ \emph {et~al.}(2015)\citenamefont {Berlin},
  \citenamefont {Gori}, \citenamefont {Lin},\ and\ \citenamefont
  {Wang}}]{Berlin:2015wwa}%
  \BibitemOpen
  \bibfield  {author} {\bibinfo {author} {\bibfnamefont {A.}~\bibnamefont
  {Berlin}}, \bibinfo {author} {\bibfnamefont {S.}~\bibnamefont {Gori}},
  \bibinfo {author} {\bibfnamefont {T.}~\bibnamefont {Lin}}, \ and\ \bibinfo
  {author} {\bibfnamefont {L.-T.}\ \bibnamefont {Wang}},\ }\href {\doibase
  10.1103/PhysRevD.92.015005} {\bibfield  {journal} {\bibinfo  {journal} {Phys.
  Rev. D}\ }\textbf {\bibinfo {volume} {92}},\ \bibinfo {pages} {015005}
  (\bibinfo {year} {2015})},\ \Eprint {http://arxiv.org/abs/1502.06000}
  {arXiv:1502.06000 [hep-ph]} \BibitemShut {NoStop}%
\bibitem [{\citenamefont {Karwin}\ \emph {et~al.}(2017)\citenamefont {Karwin},
  \citenamefont {Murgia}, \citenamefont {Tait}, \citenamefont {Porter},\ and\
  \citenamefont {Tanedo}}]{Karwin:2016tsw}%
  \BibitemOpen
  \bibfield  {author} {\bibinfo {author} {\bibfnamefont {C.}~\bibnamefont
  {Karwin}}, \bibinfo {author} {\bibfnamefont {S.}~\bibnamefont {Murgia}},
  \bibinfo {author} {\bibfnamefont {T.~M.~P.}\ \bibnamefont {Tait}}, \bibinfo
  {author} {\bibfnamefont {T.~A.}\ \bibnamefont {Porter}}, \ and\ \bibinfo
  {author} {\bibfnamefont {P.}~\bibnamefont {Tanedo}},\ }\href {\doibase
  10.1103/PhysRevD.95.103005} {\bibfield  {journal} {\bibinfo  {journal} {Phys.
  Rev. D}\ }\textbf {\bibinfo {volume} {95}},\ \bibinfo {pages} {103005}
  (\bibinfo {year} {2017})},\ \Eprint {http://arxiv.org/abs/1612.05687}
  {arXiv:1612.05687 [hep-ph]} \BibitemShut {NoStop}%
\bibitem [{\citenamefont {Abazajian}(2011)}]{Abazajian:2010zy}%
  \BibitemOpen
  \bibfield  {author} {\bibinfo {author} {\bibfnamefont {K.~N.}\ \bibnamefont
  {Abazajian}},\ }\href {\doibase 10.1088/1475-7516/2011/03/010} {\bibfield
  {journal} {\bibinfo  {journal} {JCAP}\ }\textbf {\bibinfo {volume} {03}},\
  \bibinfo {pages} {010} (\bibinfo {year} {2011})},\ \Eprint
  {http://arxiv.org/abs/1011.4275} {arXiv:1011.4275 [astro-ph.HE]} \BibitemShut
  {NoStop}%
\bibitem [{\citenamefont {Hooper}\ \emph {et~al.}(2013)\citenamefont {Hooper},
  \citenamefont {Cholis}, \citenamefont {Linden}, \citenamefont
  {Siegal-Gaskins},\ and\ \citenamefont {Slatyer}}]{Hooper:2013nhl}%
  \BibitemOpen
  \bibfield  {author} {\bibinfo {author} {\bibfnamefont {D.}~\bibnamefont
  {Hooper}}, \bibinfo {author} {\bibfnamefont {I.}~\bibnamefont {Cholis}},
  \bibinfo {author} {\bibfnamefont {T.}~\bibnamefont {Linden}}, \bibinfo
  {author} {\bibfnamefont {J.}~\bibnamefont {Siegal-Gaskins}}, \ and\ \bibinfo
  {author} {\bibfnamefont {T.}~\bibnamefont {Slatyer}},\ }\href {\doibase
  10.1103/PhysRevD.88.083009} {\bibfield  {journal} {\bibinfo  {journal} {Phys.
  Rev. D}\ }\textbf {\bibinfo {volume} {88}},\ \bibinfo {pages} {083009}
  (\bibinfo {year} {2013})},\ \Eprint {http://arxiv.org/abs/1305.0830}
  {arXiv:1305.0830 [astro-ph.HE]} \BibitemShut {NoStop}%
\bibitem [{\citenamefont {Gordon}\ and\ \citenamefont
  {Macias}(2013)}]{Gordon:2013vta}%
  \BibitemOpen
  \bibfield  {author} {\bibinfo {author} {\bibfnamefont {C.}~\bibnamefont
  {Gordon}}\ and\ \bibinfo {author} {\bibfnamefont {O.}~\bibnamefont
  {Macias}},\ }\href {\doibase 10.1103/PhysRevD.88.083521} {\bibfield
  {journal} {\bibinfo  {journal} {Phys. Rev. D}\ }\textbf {\bibinfo {volume}
  {88}},\ \bibinfo {pages} {083521} (\bibinfo {year} {2013})},\ \bibinfo {note}
  {[Erratum: Phys.Rev.D 89, 049901 (2014)]},\ \Eprint
  {http://arxiv.org/abs/1306.5725} {arXiv:1306.5725 [astro-ph.HE]} \BibitemShut
  {NoStop}%
\bibitem [{\citenamefont {Cholis}\ \emph {et~al.}(2015)\citenamefont {Cholis},
  \citenamefont {Hooper},\ and\ \citenamefont {Linden}}]{Cholis:2014lta}%
  \BibitemOpen
  \bibfield  {author} {\bibinfo {author} {\bibfnamefont {I.}~\bibnamefont
  {Cholis}}, \bibinfo {author} {\bibfnamefont {D.}~\bibnamefont {Hooper}}, \
  and\ \bibinfo {author} {\bibfnamefont {T.}~\bibnamefont {Linden}},\ }\href
  {\doibase 10.1088/1475-7516/2015/06/043} {\bibfield  {journal} {\bibinfo
  {journal} {JCAP}\ }\textbf {\bibinfo {volume} {06}},\ \bibinfo {pages} {043}
  (\bibinfo {year} {2015})},\ \Eprint {http://arxiv.org/abs/1407.5625}
  {arXiv:1407.5625 [astro-ph.HE]} \BibitemShut {NoStop}%
\bibitem [{\citenamefont {Petrovi\'c}\ \emph {et~al.}(2015)\citenamefont
  {Petrovi\'c}, \citenamefont {Serpico},\ and\ \citenamefont
  {Zaharijas}}]{Petrovic:2014xra}%
  \BibitemOpen
  \bibfield  {author} {\bibinfo {author} {\bibfnamefont {J.}~\bibnamefont
  {Petrovi\'c}}, \bibinfo {author} {\bibfnamefont {P.~D.}\ \bibnamefont
  {Serpico}}, \ and\ \bibinfo {author} {\bibfnamefont {G.}~\bibnamefont
  {Zaharijas}},\ }\href {\doibase 10.1088/1475-7516/2015/02/023} {\bibfield
  {journal} {\bibinfo  {journal} {JCAP}\ }\textbf {\bibinfo {volume} {02}},\
  \bibinfo {pages} {023} (\bibinfo {year} {2015})},\ \Eprint
  {http://arxiv.org/abs/1411.2980} {arXiv:1411.2980 [astro-ph.HE]} \BibitemShut
  {NoStop}%
\bibitem [{\citenamefont {Brandt}\ and\ \citenamefont
  {Kocsis}(2015)}]{Brandt:2015ula}%
  \BibitemOpen
  \bibfield  {author} {\bibinfo {author} {\bibfnamefont {T.~D.}\ \bibnamefont
  {Brandt}}\ and\ \bibinfo {author} {\bibfnamefont {B.}~\bibnamefont
  {Kocsis}},\ }\href {\doibase 10.1088/0004-637X/812/1/15} {\bibfield
  {journal} {\bibinfo  {journal} {Astrophys. J.}\ }\textbf {\bibinfo {volume}
  {812}},\ \bibinfo {pages} {15} (\bibinfo {year} {2015})},\ \Eprint
  {http://arxiv.org/abs/1507.05616} {arXiv:1507.05616 [astro-ph.HE]}
  \BibitemShut {NoStop}%
\bibitem [{\citenamefont {Hooper}\ and\ \citenamefont
  {Mohlabeng}(2016)}]{Hooper:2015jlu}%
  \BibitemOpen
  \bibfield  {author} {\bibinfo {author} {\bibfnamefont {D.}~\bibnamefont
  {Hooper}}\ and\ \bibinfo {author} {\bibfnamefont {G.}~\bibnamefont
  {Mohlabeng}},\ }\href {\doibase 10.1088/1475-7516/2016/03/049} {\bibfield
  {journal} {\bibinfo  {journal} {JCAP}\ }\textbf {\bibinfo {volume} {03}},\
  \bibinfo {pages} {049} (\bibinfo {year} {2016})},\ \Eprint
  {http://arxiv.org/abs/1512.04966} {arXiv:1512.04966 [astro-ph.HE]}
  \BibitemShut {NoStop}%
\bibitem [{\citenamefont {Hooper}\ and\ \citenamefont
  {Linden}(2016)}]{Hooper:2016rap}%
  \BibitemOpen
  \bibfield  {author} {\bibinfo {author} {\bibfnamefont {D.}~\bibnamefont
  {Hooper}}\ and\ \bibinfo {author} {\bibfnamefont {T.}~\bibnamefont
  {Linden}},\ }\href {\doibase 10.1088/1475-7516/2016/08/018} {\bibfield
  {journal} {\bibinfo  {journal} {JCAP}\ }\textbf {\bibinfo {volume} {08}},\
  \bibinfo {pages} {018} (\bibinfo {year} {2016})},\ \Eprint
  {http://arxiv.org/abs/1606.09250} {arXiv:1606.09250 [astro-ph.HE]}
  \BibitemShut {NoStop}%
\bibitem [{\citenamefont {Lee}\ \emph {et~al.}(2016)\citenamefont {Lee},
  \citenamefont {Lisanti}, \citenamefont {Safdi}, \citenamefont {Slatyer},\
  and\ \citenamefont {Xue}}]{Lee:2015fea}%
  \BibitemOpen
  \bibfield  {author} {\bibinfo {author} {\bibfnamefont {S.~K.}\ \bibnamefont
  {Lee}}, \bibinfo {author} {\bibfnamefont {M.}~\bibnamefont {Lisanti}},
  \bibinfo {author} {\bibfnamefont {B.~R.}\ \bibnamefont {Safdi}}, \bibinfo
  {author} {\bibfnamefont {T.~R.}\ \bibnamefont {Slatyer}}, \ and\ \bibinfo
  {author} {\bibfnamefont {W.}~\bibnamefont {Xue}},\ }\href {\doibase
  10.1103/PhysRevLett.116.051103} {\bibfield  {journal} {\bibinfo  {journal}
  {Phys. Rev. Lett.}\ }\textbf {\bibinfo {volume} {116}},\ \bibinfo {pages}
  {051103} (\bibinfo {year} {2016})},\ \Eprint
  {http://arxiv.org/abs/1506.05124} {arXiv:1506.05124 [astro-ph.HE]}
  \BibitemShut {NoStop}%
\bibitem [{\citenamefont {Bartels}\ \emph {et~al.}(2016)\citenamefont
  {Bartels}, \citenamefont {Krishnamurthy},\ and\ \citenamefont
  {Weniger}}]{Bartels:2015aea}%
  \BibitemOpen
  \bibfield  {author} {\bibinfo {author} {\bibfnamefont {R.}~\bibnamefont
  {Bartels}}, \bibinfo {author} {\bibfnamefont {S.}~\bibnamefont
  {Krishnamurthy}}, \ and\ \bibinfo {author} {\bibfnamefont {C.}~\bibnamefont
  {Weniger}},\ }\href {\doibase 10.1103/PhysRevLett.116.051102} {\bibfield
  {journal} {\bibinfo  {journal} {Phys. Rev. Lett.}\ }\textbf {\bibinfo
  {volume} {116}},\ \bibinfo {pages} {051102} (\bibinfo {year} {2016})},\
  \Eprint {http://arxiv.org/abs/1506.05104} {arXiv:1506.05104 [astro-ph.HE]}
  \BibitemShut {NoStop}%
\bibitem [{\citenamefont {Bartels}\ \emph
  {et~al.}(2018{\natexlab{a}})\citenamefont {Bartels}, \citenamefont {Storm},
  \citenamefont {Weniger},\ and\ \citenamefont {Calore}}]{Bartels:2017vsx}%
  \BibitemOpen
  \bibfield  {author} {\bibinfo {author} {\bibfnamefont {R.}~\bibnamefont
  {Bartels}}, \bibinfo {author} {\bibfnamefont {E.}~\bibnamefont {Storm}},
  \bibinfo {author} {\bibfnamefont {C.}~\bibnamefont {Weniger}}, \ and\
  \bibinfo {author} {\bibfnamefont {F.}~\bibnamefont {Calore}},\ }\href
  {\doibase 10.1038/s41550-018-0531-z} {\bibfield  {journal} {\bibinfo
  {journal} {Nature Astron.}\ }\textbf {\bibinfo {volume} {2}},\ \bibinfo
  {pages} {819} (\bibinfo {year} {2018}{\natexlab{a}})},\ \Eprint
  {http://arxiv.org/abs/1711.04778} {arXiv:1711.04778 [astro-ph.HE]}
  \BibitemShut {NoStop}%
\bibitem [{\citenamefont {Macias}\ \emph {et~al.}(2018)\citenamefont {Macias},
  \citenamefont {Gordon}, \citenamefont {Crocker}, \citenamefont {Coleman},
  \citenamefont {Paterson}, \citenamefont {Horiuchi},\ and\ \citenamefont
  {Pohl}}]{Macias:2016nev}%
  \BibitemOpen
  \bibfield  {author} {\bibinfo {author} {\bibfnamefont {O.}~\bibnamefont
  {Macias}}, \bibinfo {author} {\bibfnamefont {C.}~\bibnamefont {Gordon}},
  \bibinfo {author} {\bibfnamefont {R.~M.}\ \bibnamefont {Crocker}}, \bibinfo
  {author} {\bibfnamefont {B.}~\bibnamefont {Coleman}}, \bibinfo {author}
  {\bibfnamefont {D.}~\bibnamefont {Paterson}}, \bibinfo {author}
  {\bibfnamefont {S.}~\bibnamefont {Horiuchi}}, \ and\ \bibinfo {author}
  {\bibfnamefont {M.}~\bibnamefont {Pohl}},\ }\href {\doibase
  10.1038/s41550-018-0414-3} {\bibfield  {journal} {\bibinfo  {journal} {Nature
  Astron.}\ }\textbf {\bibinfo {volume} {2}},\ \bibinfo {pages} {387} (\bibinfo
  {year} {2018})},\ \Eprint {http://arxiv.org/abs/1611.06644} {arXiv:1611.06644
  [astro-ph.HE]} \BibitemShut {NoStop}%
\bibitem [{\citenamefont {Macias}\ \emph {et~al.}(2019)\citenamefont {Macias},
  \citenamefont {Horiuchi}, \citenamefont {Kaplinghat}, \citenamefont {Gordon},
  \citenamefont {Crocker},\ and\ \citenamefont {Nataf}}]{Macias:2019omb}%
  \BibitemOpen
  \bibfield  {author} {\bibinfo {author} {\bibfnamefont {O.}~\bibnamefont
  {Macias}}, \bibinfo {author} {\bibfnamefont {S.}~\bibnamefont {Horiuchi}},
  \bibinfo {author} {\bibfnamefont {M.}~\bibnamefont {Kaplinghat}}, \bibinfo
  {author} {\bibfnamefont {C.}~\bibnamefont {Gordon}}, \bibinfo {author}
  {\bibfnamefont {R.~M.}\ \bibnamefont {Crocker}}, \ and\ \bibinfo {author}
  {\bibfnamefont {D.~M.}\ \bibnamefont {Nataf}},\ }\href {\doibase
  10.1088/1475-7516/2019/09/042} {\bibfield  {journal} {\bibinfo  {journal}
  {JCAP}\ }\textbf {\bibinfo {volume} {09}},\ \bibinfo {pages} {042} (\bibinfo
  {year} {2019})},\ \Eprint {http://arxiv.org/abs/1901.03822} {arXiv:1901.03822
  [astro-ph.HE]} \BibitemShut {NoStop}%
\bibitem [{\citenamefont {Pohl}\ \emph {et~al.}(2022)\citenamefont {Pohl},
  \citenamefont {Macias}, \citenamefont {Coleman},\ and\ \citenamefont
  {Gordon}}]{Pohl:2022nnd}%
  \BibitemOpen
  \bibfield  {author} {\bibinfo {author} {\bibfnamefont {M.}~\bibnamefont
  {Pohl}}, \bibinfo {author} {\bibfnamefont {O.}~\bibnamefont {Macias}},
  \bibinfo {author} {\bibfnamefont {P.}~\bibnamefont {Coleman}}, \ and\
  \bibinfo {author} {\bibfnamefont {C.}~\bibnamefont {Gordon}},\ }\href
  {\doibase 10.3847/1538-4357/ac6032} {\bibfield  {journal} {\bibinfo
  {journal} {Astrophys. J.}\ }\textbf {\bibinfo {volume} {929}},\ \bibinfo
  {pages} {136} (\bibinfo {year} {2022})},\ \Eprint
  {http://arxiv.org/abs/2203.11626} {arXiv:2203.11626 [astro-ph.HE]}
  \BibitemShut {NoStop}%
\bibitem [{\citenamefont {Hooper}(2022)}]{Hooper:2022bec}%
  \BibitemOpen
  \bibfield  {author} {\bibinfo {author} {\bibfnamefont {D.}~\bibnamefont
  {Hooper}},\ }in\ \href@noop {} {\emph {\bibinfo {booktitle} {{14th
  International Workshop on the Identification of Dark Matter 2022}}}}\
  (\bibinfo {year} {2022})\ \Eprint {http://arxiv.org/abs/2209.14370}
  {arXiv:2209.14370 [astro-ph.HE]} \BibitemShut {NoStop}%
\bibitem [{\citenamefont {Leane}\ and\ \citenamefont
  {Slatyer}(2019)}]{Leane:2019xiy}%
  \BibitemOpen
  \bibfield  {author} {\bibinfo {author} {\bibfnamefont {R.~K.}\ \bibnamefont
  {Leane}}\ and\ \bibinfo {author} {\bibfnamefont {T.~R.}\ \bibnamefont
  {Slatyer}},\ }\href {\doibase 10.1103/PhysRevLett.123.241101} {\bibfield
  {journal} {\bibinfo  {journal} {Phys. Rev. Lett.}\ }\textbf {\bibinfo
  {volume} {123}},\ \bibinfo {pages} {241101} (\bibinfo {year} {2019})},\
  \Eprint {http://arxiv.org/abs/1904.08430} {arXiv:1904.08430 [astro-ph.HE]}
  \BibitemShut {NoStop}%
\bibitem [{\citenamefont {Leane}\ and\ \citenamefont
  {Slatyer}(2020{\natexlab{a}})}]{Leane:2020pfc}%
  \BibitemOpen
  \bibfield  {author} {\bibinfo {author} {\bibfnamefont {R.~K.}\ \bibnamefont
  {Leane}}\ and\ \bibinfo {author} {\bibfnamefont {T.~R.}\ \bibnamefont
  {Slatyer}},\ }\href {\doibase 10.1103/PhysRevD.102.063019} {\bibfield
  {journal} {\bibinfo  {journal} {Phys. Rev. D}\ }\textbf {\bibinfo {volume}
  {102}},\ \bibinfo {pages} {063019} (\bibinfo {year} {2020}{\natexlab{a}})},\
  \Eprint {http://arxiv.org/abs/2002.12371} {arXiv:2002.12371 [astro-ph.HE]}
  \BibitemShut {NoStop}%
\bibitem [{\citenamefont {Leane}\ and\ \citenamefont
  {Slatyer}(2020{\natexlab{b}})}]{Leane:2020nmi}%
  \BibitemOpen
  \bibfield  {author} {\bibinfo {author} {\bibfnamefont {R.~K.}\ \bibnamefont
  {Leane}}\ and\ \bibinfo {author} {\bibfnamefont {T.~R.}\ \bibnamefont
  {Slatyer}},\ }\href {\doibase 10.1103/PhysRevLett.125.121105} {\bibfield
  {journal} {\bibinfo  {journal} {Phys. Rev. Lett.}\ }\textbf {\bibinfo
  {volume} {125}},\ \bibinfo {pages} {121105} (\bibinfo {year}
  {2020}{\natexlab{b}})},\ \Eprint {http://arxiv.org/abs/2002.12370}
  {arXiv:2002.12370 [astro-ph.HE]} \BibitemShut {NoStop}%
\bibitem [{\citenamefont {Zhong}\ \emph {et~al.}(2020)\citenamefont {Zhong},
  \citenamefont {McDermott}, \citenamefont {Cholis},\ and\ \citenamefont
  {Fox}}]{Zhong:2019ycb}%
  \BibitemOpen
  \bibfield  {author} {\bibinfo {author} {\bibfnamefont {Y.-M.}\ \bibnamefont
  {Zhong}}, \bibinfo {author} {\bibfnamefont {S.~D.}\ \bibnamefont
  {McDermott}}, \bibinfo {author} {\bibfnamefont {I.}~\bibnamefont {Cholis}}, \
  and\ \bibinfo {author} {\bibfnamefont {P.~J.}\ \bibnamefont {Fox}},\ }\href
  {\doibase 10.1103/PhysRevLett.124.231103} {\bibfield  {journal} {\bibinfo
  {journal} {Phys. Rev. Lett.}\ }\textbf {\bibinfo {volume} {124}},\ \bibinfo
  {pages} {231103} (\bibinfo {year} {2020})},\ \Eprint
  {http://arxiv.org/abs/1911.12369} {arXiv:1911.12369 [astro-ph.HE]}
  \BibitemShut {NoStop}%
\bibitem [{\citenamefont {McDermott}\ \emph {et~al.}(2022)\citenamefont
  {McDermott}, \citenamefont {Zhong},\ and\ \citenamefont
  {Cholis}}]{McDermott:2022zmq}%
  \BibitemOpen
  \bibfield  {author} {\bibinfo {author} {\bibfnamefont {S.~D.}\ \bibnamefont
  {McDermott}}, \bibinfo {author} {\bibfnamefont {Y.-M.}\ \bibnamefont
  {Zhong}}, \ and\ \bibinfo {author} {\bibfnamefont {I.}~\bibnamefont
  {Cholis}},\ }\href@noop {} {\  (\bibinfo {year} {2022})},\ \Eprint
  {http://arxiv.org/abs/2209.00006} {arXiv:2209.00006 [astro-ph.HE]}
  \BibitemShut {NoStop}%
\bibitem [{\citenamefont {Bartels}\ \emph
  {et~al.}(2018{\natexlab{b}})\citenamefont {Bartels}, \citenamefont {Hooper},
  \citenamefont {Linden}, \citenamefont {Mishra-Sharma}, \citenamefont {Rodd},
  \citenamefont {Safdi},\ and\ \citenamefont {Slatyer}}]{Bartels:2017xba}%
  \BibitemOpen
  \bibfield  {author} {\bibinfo {author} {\bibfnamefont {R.}~\bibnamefont
  {Bartels}}, \bibinfo {author} {\bibfnamefont {D.}~\bibnamefont {Hooper}},
  \bibinfo {author} {\bibfnamefont {T.}~\bibnamefont {Linden}}, \bibinfo
  {author} {\bibfnamefont {S.}~\bibnamefont {Mishra-Sharma}}, \bibinfo {author}
  {\bibfnamefont {N.~L.}\ \bibnamefont {Rodd}}, \bibinfo {author}
  {\bibfnamefont {B.~R.}\ \bibnamefont {Safdi}}, \ and\ \bibinfo {author}
  {\bibfnamefont {T.~R.}\ \bibnamefont {Slatyer}},\ }\href {\doibase
  10.1016/j.dark.2018.04.004} {\bibfield  {journal} {\bibinfo  {journal} {Phys.
  Dark Univ.}\ }\textbf {\bibinfo {volume} {20}},\ \bibinfo {pages} {88}
  (\bibinfo {year} {2018}{\natexlab{b}})},\ \Eprint
  {http://arxiv.org/abs/1710.10266} {arXiv:1710.10266 [astro-ph.HE]}
  \BibitemShut {NoStop}%
\bibitem [{\citenamefont {Dinsmore}\ and\ \citenamefont
  {Slatyer}(2022)}]{Dinsmore:2021nip}%
  \BibitemOpen
  \bibfield  {author} {\bibinfo {author} {\bibfnamefont {J.~T.}\ \bibnamefont
  {Dinsmore}}\ and\ \bibinfo {author} {\bibfnamefont {T.~R.}\ \bibnamefont
  {Slatyer}},\ }\href {\doibase 10.1088/1475-7516/2022/06/025} {\bibfield
  {journal} {\bibinfo  {journal} {JCAP}\ }\textbf {\bibinfo {volume} {06}},\
  \bibinfo {pages} {025} (\bibinfo {year} {2022})},\ \Eprint
  {http://arxiv.org/abs/2112.09699} {arXiv:2112.09699 [astro-ph.HE]}
  \BibitemShut {NoStop}%
\bibitem [{\citenamefont {Calore}\ \emph {et~al.}(2016)\citenamefont {Calore},
  \citenamefont {Di~Mauro}, \citenamefont {Donato}, \citenamefont {Hessels},\
  and\ \citenamefont {Weniger}}]{Calore:2015bsx}%
  \BibitemOpen
  \bibfield  {author} {\bibinfo {author} {\bibfnamefont {F.}~\bibnamefont
  {Calore}}, \bibinfo {author} {\bibfnamefont {M.}~\bibnamefont {Di~Mauro}},
  \bibinfo {author} {\bibfnamefont {F.}~\bibnamefont {Donato}}, \bibinfo
  {author} {\bibfnamefont {J.~W.~T.}\ \bibnamefont {Hessels}}, \ and\ \bibinfo
  {author} {\bibfnamefont {C.}~\bibnamefont {Weniger}},\ }\href {\doibase
  10.3847/0004-637X/827/2/143} {\bibfield  {journal} {\bibinfo  {journal}
  {Astrophys. J.}\ }\textbf {\bibinfo {volume} {827}},\ \bibinfo {pages} {143}
  (\bibinfo {year} {2016})},\ \Eprint {http://arxiv.org/abs/1512.06825}
  {arXiv:1512.06825 [astro-ph.HE]} \BibitemShut {NoStop}%
\bibitem [{\citenamefont {Haggard}\ \emph {et~al.}(2017)\citenamefont
  {Haggard}, \citenamefont {Heinke}, \citenamefont {Hooper},\ and\
  \citenamefont {Linden}}]{Haggard:2017lyq}%
  \BibitemOpen
  \bibfield  {author} {\bibinfo {author} {\bibfnamefont {D.}~\bibnamefont
  {Haggard}}, \bibinfo {author} {\bibfnamefont {C.}~\bibnamefont {Heinke}},
  \bibinfo {author} {\bibfnamefont {D.}~\bibnamefont {Hooper}}, \ and\ \bibinfo
  {author} {\bibfnamefont {T.}~\bibnamefont {Linden}},\ }\href {\doibase
  10.1088/1475-7516/2017/05/056} {\bibfield  {journal} {\bibinfo  {journal}
  {JCAP}\ }\textbf {\bibinfo {volume} {05}},\ \bibinfo {pages} {056} (\bibinfo
  {year} {2017})},\ \Eprint {http://arxiv.org/abs/1701.02726} {arXiv:1701.02726
  [astro-ph.HE]} \BibitemShut {NoStop}%
\bibitem [{\citenamefont {Hooper}\ \emph {et~al.}(2017)\citenamefont {Hooper},
  \citenamefont {Cholis}, \citenamefont {Linden},\ and\ \citenamefont
  {Fang}}]{Hooper:2017gtd}%
  \BibitemOpen
  \bibfield  {author} {\bibinfo {author} {\bibfnamefont {D.}~\bibnamefont
  {Hooper}}, \bibinfo {author} {\bibfnamefont {I.}~\bibnamefont {Cholis}},
  \bibinfo {author} {\bibfnamefont {T.}~\bibnamefont {Linden}}, \ and\ \bibinfo
  {author} {\bibfnamefont {K.}~\bibnamefont {Fang}},\ }\href {\doibase
  10.1103/PhysRevD.96.103013} {\bibfield  {journal} {\bibinfo  {journal} {Phys.
  Rev. D}\ }\textbf {\bibinfo {volume} {96}},\ \bibinfo {pages} {103013}
  (\bibinfo {year} {2017})},\ \Eprint {http://arxiv.org/abs/1702.08436}
  {arXiv:1702.08436 [astro-ph.HE]} \BibitemShut {NoStop}%
\bibitem [{\citenamefont {Linden}\ \emph {et~al.}(2017)\citenamefont {Linden},
  \citenamefont {Auchettl}, \citenamefont {Bramante}, \citenamefont {Cholis},
  \citenamefont {Fang}, \citenamefont {Hooper}, \citenamefont {Karwal},\ and\
  \citenamefont {Li}}]{Linden:2017vvb}%
  \BibitemOpen
  \bibfield  {author} {\bibinfo {author} {\bibfnamefont {T.}~\bibnamefont
  {Linden}}, \bibinfo {author} {\bibfnamefont {K.}~\bibnamefont {Auchettl}},
  \bibinfo {author} {\bibfnamefont {J.}~\bibnamefont {Bramante}}, \bibinfo
  {author} {\bibfnamefont {I.}~\bibnamefont {Cholis}}, \bibinfo {author}
  {\bibfnamefont {K.}~\bibnamefont {Fang}}, \bibinfo {author} {\bibfnamefont
  {D.}~\bibnamefont {Hooper}}, \bibinfo {author} {\bibfnamefont
  {T.}~\bibnamefont {Karwal}}, \ and\ \bibinfo {author} {\bibfnamefont {S.~W.}\
  \bibnamefont {Li}},\ }\href {\doibase 10.1103/PhysRevD.96.103016} {\bibfield
  {journal} {\bibinfo  {journal} {Phys. Rev. D}\ }\textbf {\bibinfo {volume}
  {96}},\ \bibinfo {pages} {103016} (\bibinfo {year} {2017})},\ \Eprint
  {http://arxiv.org/abs/1703.09704} {arXiv:1703.09704 [astro-ph.HE]}
  \BibitemShut {NoStop}%
\bibitem [{\citenamefont {Albert}\ \emph {et~al.}(2021)\citenamefont {Albert}
  \emph {et~al.}}]{HAWC:2021dtl}%
  \BibitemOpen
  \bibfield  {author} {\bibinfo {author} {\bibfnamefont {A.}~\bibnamefont
  {Albert}} \emph {et~al.} (\bibinfo {collaboration} {HAWC}),\ }\href {\doibase
  10.3847/2041-8213/abf4dc} {\bibfield  {journal} {\bibinfo  {journal}
  {Astrophys. J. Lett.}\ }\textbf {\bibinfo {volume} {911}},\ \bibinfo {pages}
  {L27} (\bibinfo {year} {2021})},\ \Eprint {http://arxiv.org/abs/2101.07895}
  {arXiv:2101.07895 [astro-ph.HE]} \BibitemShut {NoStop}%
\bibitem [{\citenamefont {Abeysekara}\ \emph {et~al.}(2020)\citenamefont
  {Abeysekara} \emph {et~al.}}]{HAWC:2019tcx}%
  \BibitemOpen
  \bibfield  {author} {\bibinfo {author} {\bibfnamefont {A.~U.}\ \bibnamefont
  {Abeysekara}} \emph {et~al.} (\bibinfo {collaboration} {HAWC}),\ }\href
  {\doibase 10.1103/PhysRevLett.124.021102} {\bibfield  {journal} {\bibinfo
  {journal} {Phys. Rev. Lett.}\ }\textbf {\bibinfo {volume} {124}},\ \bibinfo
  {pages} {021102} (\bibinfo {year} {2020})},\ \Eprint
  {http://arxiv.org/abs/1909.08609} {arXiv:1909.08609 [astro-ph.HE]}
  \BibitemShut {NoStop}%
\bibitem [{\citenamefont {Hooper}\ and\ \citenamefont
  {Linden}(2022)}]{Hooper:2021kyp}%
  \BibitemOpen
  \bibfield  {author} {\bibinfo {author} {\bibfnamefont {D.}~\bibnamefont
  {Hooper}}\ and\ \bibinfo {author} {\bibfnamefont {T.}~\bibnamefont
  {Linden}},\ }\href {\doibase 10.1103/PhysRevD.105.103013} {\bibfield
  {journal} {\bibinfo  {journal} {Phys. Rev. D}\ }\textbf {\bibinfo {volume}
  {105}},\ \bibinfo {pages} {103013} (\bibinfo {year} {2022})},\ \Eprint
  {http://arxiv.org/abs/2104.00014} {arXiv:2104.00014 [astro-ph.HE]}
  \BibitemShut {NoStop}%
\bibitem [{\citenamefont {Hooper}\ and\ \citenamefont
  {Linden}(2018)}]{Hooper:2018fih}%
  \BibitemOpen
  \bibfield  {author} {\bibinfo {author} {\bibfnamefont {D.}~\bibnamefont
  {Hooper}}\ and\ \bibinfo {author} {\bibfnamefont {T.}~\bibnamefont
  {Linden}},\ }\href {\doibase 10.1103/PhysRevD.98.043005} {\bibfield
  {journal} {\bibinfo  {journal} {Phys. Rev. D}\ }\textbf {\bibinfo {volume}
  {98}},\ \bibinfo {pages} {043005} (\bibinfo {year} {2018})},\ \Eprint
  {http://arxiv.org/abs/1803.08046} {arXiv:1803.08046 [astro-ph.HE]}
  \BibitemShut {NoStop}%
\bibitem [{\citenamefont {Albert}\ \emph {et~al.}(2020)\citenamefont {Albert}
  \emph {et~al.}}]{HAWC:2020hrt}%
  \BibitemOpen
  \bibfield  {author} {\bibinfo {author} {\bibfnamefont {A.}~\bibnamefont
  {Albert}} \emph {et~al.} (\bibinfo {collaboration} {HAWC}),\ }\href {\doibase
  10.3847/1538-4357/abc2d8} {\bibfield  {journal} {\bibinfo  {journal}
  {Astrophys. J.}\ }\textbf {\bibinfo {volume} {905}},\ \bibinfo {pages} {76}
  (\bibinfo {year} {2020})},\ \Eprint {http://arxiv.org/abs/2007.08582}
  {arXiv:2007.08582 [astro-ph.HE]} \BibitemShut {NoStop}%
\bibitem [{\citenamefont {Abeysekara}\ \emph
  {et~al.}(2017{\natexlab{a}})\citenamefont {Abeysekara} \emph
  {et~al.}}]{Abeysekara:2017hyn}%
  \BibitemOpen
  \bibfield  {author} {\bibinfo {author} {\bibfnamefont {A.~U.}\ \bibnamefont
  {Abeysekara}} \emph {et~al.},\ }\href {\doibase 10.3847/1538-4357/aa7556}
  {\bibfield  {journal} {\bibinfo  {journal} {Astrophys. J.}\ }\textbf
  {\bibinfo {volume} {843}},\ \bibinfo {pages} {40} (\bibinfo {year}
  {2017}{\natexlab{a}})},\ \Eprint {http://arxiv.org/abs/1702.02992}
  {arXiv:1702.02992 [astro-ph.HE]} \BibitemShut {NoStop}%
\bibitem [{\citenamefont {Abeysekara}\ \emph
  {et~al.}(2017{\natexlab{b}})\citenamefont {Abeysekara} \emph
  {et~al.}}]{Abeysekara_2017_2}%
  \BibitemOpen
  \bibfield  {author} {\bibinfo {author} {\bibfnamefont {A.~U.}\ \bibnamefont
  {Abeysekara}} \emph {et~al.} (\bibinfo {collaboration} {HAWC}),\ }\href
  {\doibase 10.1126/science.aan4880} {\bibfield  {journal} {\bibinfo  {journal}
  {Science}\ }\textbf {\bibinfo {volume} {358}},\ \bibinfo {pages} {911}
  (\bibinfo {year} {2017}{\natexlab{b}})},\ \Eprint
  {http://arxiv.org/abs/1711.06223} {arXiv:1711.06223 [astro-ph.HE]}
  \BibitemShut {NoStop}%
\bibitem [{\citenamefont {Abdo}\ \emph {et~al.}(2009)\citenamefont {Abdo} \emph
  {et~al.}}]{Abdo:2009ku}%
  \BibitemOpen
  \bibfield  {author} {\bibinfo {author} {\bibfnamefont {A.~A.}\ \bibnamefont
  {Abdo}} \emph {et~al.},\ }\href {\doibase 10.1088/0004-637X/700/2/L127}
  {\bibfield  {journal} {\bibinfo  {journal} {Astrophys. J. Lett.}\ }\textbf
  {\bibinfo {volume} {700}},\ \bibinfo {pages} {L127} (\bibinfo {year}
  {2009})},\ \bibinfo {note} {[Erratum: Astrophys.J.Lett. 703, L185 (2009),
  Erratum: Astrophys.J. 703, L185 (2009)]},\ \Eprint
  {http://arxiv.org/abs/0904.1018} {arXiv:0904.1018 [astro-ph.HE]} \BibitemShut
  {NoStop}%
\bibitem [{\citenamefont {Sudoh}\ \emph {et~al.}(2021)\citenamefont {Sudoh},
  \citenamefont {Linden},\ and\ \citenamefont {Hooper}}]{Sudoh:2021avj}%
  \BibitemOpen
  \bibfield  {author} {\bibinfo {author} {\bibfnamefont {T.}~\bibnamefont
  {Sudoh}}, \bibinfo {author} {\bibfnamefont {T.}~\bibnamefont {Linden}}, \
  and\ \bibinfo {author} {\bibfnamefont {D.}~\bibnamefont {Hooper}},\ }\href
  {\doibase 10.1088/1475-7516/2021/08/010} {\bibfield  {journal} {\bibinfo
  {journal} {JCAP}\ }\textbf {\bibinfo {volume} {08}},\ \bibinfo {pages} {010}
  (\bibinfo {year} {2021})},\ \Eprint {http://arxiv.org/abs/2101.11026}
  {arXiv:2101.11026 [astro-ph.HE]} \BibitemShut {NoStop}%
\bibitem [{\citenamefont {{Linden}}\ and\ \citenamefont
  {{Buckman}}(2018)}]{Linden_2018}%
  \BibitemOpen
  \bibfield  {author} {\bibinfo {author} {\bibfnamefont {T.}~\bibnamefont
  {{Linden}}}\ and\ \bibinfo {author} {\bibfnamefont {B.~J.}\ \bibnamefont
  {{Buckman}}},\ }\href {\doibase 10.1103/PhysRevLett.120.121101} {\bibfield
  {journal} {\bibinfo  {journal} {\prl}\ }\textbf {\bibinfo {volume} {120}},\
  \bibinfo {eid} {121101} (\bibinfo {year} {2018})},\ \Eprint
  {http://arxiv.org/abs/1707.01905} {arXiv:1707.01905 [astro-ph.HE]}
  \BibitemShut {NoStop}%
\bibitem [{\citenamefont {Xu}\ and\ \citenamefont {Hooper}(2022)}]{Xu:2021ncy}%
  \BibitemOpen
  \bibfield  {author} {\bibinfo {author} {\bibfnamefont {F.}~\bibnamefont
  {Xu}}\ and\ \bibinfo {author} {\bibfnamefont {D.}~\bibnamefont {Hooper}},\
  }\href {\doibase 10.1103/PhysRevD.106.023005} {\bibfield  {journal} {\bibinfo
   {journal} {Phys. Rev. D}\ }\textbf {\bibinfo {volume} {106}},\ \bibinfo
  {pages} {023005} (\bibinfo {year} {2022})},\ \Eprint
  {http://arxiv.org/abs/2111.03646} {arXiv:2111.03646 [astro-ph.HE]}
  \BibitemShut {NoStop}%
\bibitem [{\citenamefont {Venter}\ \emph {et~al.}(2015)\citenamefont {Venter},
  \citenamefont {Kopp}, \citenamefont {Harding}, \citenamefont {Gonthier},\
  and\ \citenamefont {B\"usching}}]{Venter:2015gga}%
  \BibitemOpen
  \bibfield  {author} {\bibinfo {author} {\bibfnamefont {C.}~\bibnamefont
  {Venter}}, \bibinfo {author} {\bibfnamefont {A.}~\bibnamefont {Kopp}},
  \bibinfo {author} {\bibfnamefont {A.~K.}\ \bibnamefont {Harding}}, \bibinfo
  {author} {\bibfnamefont {P.~L.}\ \bibnamefont {Gonthier}}, \ and\ \bibinfo
  {author} {\bibfnamefont {I.}~\bibnamefont {B\"usching}},\ }\href {\doibase
  10.1088/0004-637X/807/2/130} {\bibfield  {journal} {\bibinfo  {journal}
  {Astrophys. J.}\ }\textbf {\bibinfo {volume} {807}},\ \bibinfo {pages} {130}
  (\bibinfo {year} {2015})},\ \Eprint {http://arxiv.org/abs/1506.01211}
  {arXiv:1506.01211 [astro-ph.HE]} \BibitemShut {NoStop}%
\bibitem [{\citenamefont {Bednarek}\ \emph {et~al.}(2016)\citenamefont
  {Bednarek}, \citenamefont {Sitarek},\ and\ \citenamefont
  {Sobczak}}]{Bednarek:2016gpp}%
  \BibitemOpen
  \bibfield  {author} {\bibinfo {author} {\bibfnamefont {W.}~\bibnamefont
  {Bednarek}}, \bibinfo {author} {\bibfnamefont {J.}~\bibnamefont {Sitarek}}, \
  and\ \bibinfo {author} {\bibfnamefont {T.}~\bibnamefont {Sobczak}},\ }\href
  {\doibase 10.1093/mnras/stw367} {\bibfield  {journal} {\bibinfo  {journal}
  {Mon. Not. Roy. Astron. Soc.}\ }\textbf {\bibinfo {volume} {458}},\ \bibinfo
  {pages} {1083} (\bibinfo {year} {2016})},\ \Eprint
  {http://arxiv.org/abs/1602.03629} {arXiv:1602.03629 [astro-ph.HE]}
  \BibitemShut {NoStop}%
\bibitem [{\citenamefont {Venter}\ \emph {et~al.}(2016)\citenamefont {Venter},
  \citenamefont {Kopp}, \citenamefont {Harding}, \citenamefont {Gonthier},\
  and\ \citenamefont {B\"usching}}]{Venter:2015oza}%
  \BibitemOpen
  \bibfield  {author} {\bibinfo {author} {\bibfnamefont {C.}~\bibnamefont
  {Venter}}, \bibinfo {author} {\bibfnamefont {A.}~\bibnamefont {Kopp}},
  \bibinfo {author} {\bibfnamefont {A.~K.}\ \bibnamefont {Harding}}, \bibinfo
  {author} {\bibfnamefont {P.~L.}\ \bibnamefont {Gonthier}}, \ and\ \bibinfo
  {author} {\bibfnamefont {I.}~\bibnamefont {B\"usching}},\ }\href {\doibase
  10.22323/1.236.0462} {\bibfield  {journal} {\bibinfo  {journal} {PoS}\
  }\textbf {\bibinfo {volume} {ICRC2015}},\ \bibinfo {pages} {462} (\bibinfo
  {year} {2016})},\ \Eprint {http://arxiv.org/abs/1508.04676} {arXiv:1508.04676
  [astro-ph.HE]} \BibitemShut {NoStop}%
\bibitem [{\citenamefont {{Sironi}}\ and\ \citenamefont
  {{Spitkovsky}}(2011)}]{2011ApJ...741...39S}%
  \BibitemOpen
  \bibfield  {author} {\bibinfo {author} {\bibfnamefont {L.}~\bibnamefont
  {{Sironi}}}\ and\ \bibinfo {author} {\bibfnamefont {A.}~\bibnamefont
  {{Spitkovsky}}},\ }\href {\doibase 10.1088/0004-637X/741/1/39} {\bibfield
  {journal} {\bibinfo  {journal} {\apj}\ }\textbf {\bibinfo {volume} {741}},\
  \bibinfo {eid} {39} (\bibinfo {year} {2011})},\ \Eprint
  {http://arxiv.org/abs/1107.0977} {arXiv:1107.0977 [astro-ph.HE]} \BibitemShut
  {NoStop}%
\bibitem [{\citenamefont {Gaensler}\ and\ \citenamefont
  {Slane}(2006)}]{Gaensler:2006ua}%
  \BibitemOpen
  \bibfield  {author} {\bibinfo {author} {\bibfnamefont {B.~M.}\ \bibnamefont
  {Gaensler}}\ and\ \bibinfo {author} {\bibfnamefont {P.~O.}\ \bibnamefont
  {Slane}},\ }\href {\doibase 10.1146/annurev.astro.44.051905.092528}
  {\bibfield  {journal} {\bibinfo  {journal} {Ann. Rev. Astron. Astrophys.}\
  }\textbf {\bibinfo {volume} {44}},\ \bibinfo {pages} {17} (\bibinfo {year}
  {2006})},\ \Eprint {http://arxiv.org/abs/astro-ph/0601081}
  {arXiv:astro-ph/0601081} \BibitemShut {NoStop}%
\bibitem [{\citenamefont {BLUMENTHAL}\ and\ \citenamefont
  {GOULD}(1970)}]{RevModPhys.42.237}%
  \BibitemOpen
  \bibfield  {author} {\bibinfo {author} {\bibfnamefont {G.~R.}\ \bibnamefont
  {BLUMENTHAL}}\ and\ \bibinfo {author} {\bibfnamefont {R.~J.}\ \bibnamefont
  {GOULD}},\ }\href {\doibase 10.1103/RevModPhys.42.237} {\bibfield  {journal}
  {\bibinfo  {journal} {Rev. Mod. Phys.}\ }\textbf {\bibinfo {volume} {42}},\
  \bibinfo {pages} {237} (\bibinfo {year} {1970})}\BibitemShut {NoStop}%
\bibitem [{\citenamefont {{Schlickeiser}}\ and\ \citenamefont
  {{Ruppel}}(2010)}]{Schlickeiser_2010}%
  \BibitemOpen
  \bibfield  {author} {\bibinfo {author} {\bibfnamefont {R.}~\bibnamefont
  {{Schlickeiser}}}\ and\ \bibinfo {author} {\bibfnamefont {J.}~\bibnamefont
  {{Ruppel}}},\ }\href {\doibase 10.1088/1367-2630/12/3/033044} {\bibfield
  {journal} {\bibinfo  {journal} {New Journal of Physics}\ }\textbf {\bibinfo
  {volume} {12}},\ \bibinfo {eid} {033044} (\bibinfo {year} {2010})},\ \Eprint
  {http://arxiv.org/abs/0908.2183} {arXiv:0908.2183 [astro-ph.HE]} \BibitemShut
  {NoStop}%
\bibitem [{\citenamefont {Aharonian}\ and\ \citenamefont
  {Atoyan}(1981)}]{Aharonian:1981spy}%
  \BibitemOpen
  \bibfield  {author} {\bibinfo {author} {\bibfnamefont {F.~A.}\ \bibnamefont
  {Aharonian}}\ and\ \bibinfo {author} {\bibfnamefont {A.~M.}\ \bibnamefont
  {Atoyan}},\ }\href {\doibase 10.1016/0370-2693(81)91130-8} {\bibfield
  {journal} {\bibinfo  {journal} {Phys. Lett. B}\ }\textbf {\bibinfo {volume}
  {99}},\ \bibinfo {pages} {301} (\bibinfo {year} {1981})}\BibitemShut
  {NoStop}%
\bibitem [{\citenamefont {{Vladimirov}}\ \emph {et~al.}(2011)\citenamefont
  {{Vladimirov}}, \citenamefont {{Digel}}, \citenamefont {{J{\'o}hannesson}},
  \citenamefont {{Michelson}}, \citenamefont {{Moskalenko}}, \citenamefont
  {{Nolan}}, \citenamefont {{Orlando}}, \citenamefont {{Porter}},\ and\
  \citenamefont {{Strong}}}]{2011CoPhC.182.1156V}%
  \BibitemOpen
  \bibfield  {author} {\bibinfo {author} {\bibfnamefont {A.~E.}\ \bibnamefont
  {{Vladimirov}}}, \bibinfo {author} {\bibfnamefont {S.~W.}\ \bibnamefont
  {{Digel}}}, \bibinfo {author} {\bibfnamefont {G.}~\bibnamefont
  {{J{\'o}hannesson}}}, \bibinfo {author} {\bibfnamefont {P.~F.}\ \bibnamefont
  {{Michelson}}}, \bibinfo {author} {\bibfnamefont {I.~V.}\ \bibnamefont
  {{Moskalenko}}}, \bibinfo {author} {\bibfnamefont {P.~L.}\ \bibnamefont
  {{Nolan}}}, \bibinfo {author} {\bibfnamefont {E.}~\bibnamefont {{Orlando}}},
  \bibinfo {author} {\bibfnamefont {T.~A.}\ \bibnamefont {{Porter}}}, \ and\
  \bibinfo {author} {\bibfnamefont {A.~W.}\ \bibnamefont {{Strong}}},\ }\href
  {\doibase 10.1016/j.cpc.2011.01.017} {\bibfield  {journal} {\bibinfo
  {journal} {Computer Physics Communications}\ }\textbf {\bibinfo {volume}
  {182}},\ \bibinfo {pages} {1156} (\bibinfo {year} {2011})},\ \Eprint
  {http://arxiv.org/abs/1008.3642} {arXiv:1008.3642 [astro-ph.HE]} \BibitemShut
  {NoStop}%
\bibitem [{\citenamefont {{G{\'o}rski}}\ \emph {et~al.}(2005)\citenamefont
  {{G{\'o}rski}}, \citenamefont {{Hivon}}, \citenamefont {{Banday}},
  \citenamefont {{Wandelt}}, \citenamefont {{Hansen}}, \citenamefont
  {{Reinecke}},\ and\ \citenamefont {{Bartelmann}}}]{2005ApJ...622..759G}%
  \BibitemOpen
  \bibfield  {author} {\bibinfo {author} {\bibfnamefont {K.~M.}\ \bibnamefont
  {{G{\'o}rski}}}, \bibinfo {author} {\bibfnamefont {E.}~\bibnamefont
  {{Hivon}}}, \bibinfo {author} {\bibfnamefont {A.~J.}\ \bibnamefont
  {{Banday}}}, \bibinfo {author} {\bibfnamefont {B.~D.}\ \bibnamefont
  {{Wandelt}}}, \bibinfo {author} {\bibfnamefont {F.~K.}\ \bibnamefont
  {{Hansen}}}, \bibinfo {author} {\bibfnamefont {M.}~\bibnamefont
  {{Reinecke}}}, \ and\ \bibinfo {author} {\bibfnamefont {M.}~\bibnamefont
  {{Bartelmann}}},\ }\href {\doibase 10.1086/427976} {\bibfield  {journal}
  {\bibinfo  {journal} {\apj}\ }\textbf {\bibinfo {volume} {622}},\ \bibinfo
  {pages} {759} (\bibinfo {year} {2005})},\ \Eprint
  {http://arxiv.org/abs/astro-ph/0409513} {arXiv:astro-ph/0409513 [astro-ph]}
  \BibitemShut {NoStop}%
\bibitem [{\citenamefont {Ackermann}\ \emph {et~al.}(2014)\citenamefont
  {Ackermann} \emph {et~al.}}]{Fermi-LAT:2014sfa}%
  \BibitemOpen
  \bibfield  {author} {\bibinfo {author} {\bibfnamefont {M.}~\bibnamefont
  {Ackermann}} \emph {et~al.} (\bibinfo {collaboration} {Fermi-LAT}),\ }\href
  {\doibase 10.1088/0004-637X/793/1/64} {\bibfield  {journal} {\bibinfo
  {journal} {Astrophys. J.}\ }\textbf {\bibinfo {volume} {793}},\ \bibinfo
  {pages} {64} (\bibinfo {year} {2014})},\ \Eprint
  {http://arxiv.org/abs/1407.7905} {arXiv:1407.7905 [astro-ph.HE]} \BibitemShut
  {NoStop}%
\bibitem [{\citenamefont {Abdollahi}\ \emph {et~al.}(2020)\citenamefont
  {Abdollahi} \emph {et~al.}}]{Fermi-LAT:2019yla}%
  \BibitemOpen
  \bibfield  {author} {\bibinfo {author} {\bibfnamefont {S.}~\bibnamefont
  {Abdollahi}} \emph {et~al.} (\bibinfo {collaboration} {Fermi-LAT}),\ }\href
  {\doibase 10.3847/1538-4365/ab6bcb} {\bibfield  {journal} {\bibinfo
  {journal} {Astrophys. J. Suppl.}\ }\textbf {\bibinfo {volume} {247}},\
  \bibinfo {pages} {33} (\bibinfo {year} {2020})},\ \Eprint
  {http://arxiv.org/abs/1902.10045} {arXiv:1902.10045 [astro-ph.HE]}
  \BibitemShut {NoStop}%
\bibitem [{\citenamefont {Abdo}\ \emph {et~al.}(2013)\citenamefont {Abdo} \emph
  {et~al.}}]{Fermi-LAT:2013svs}%
  \BibitemOpen
  \bibfield  {author} {\bibinfo {author} {\bibfnamefont {A.~A.}\ \bibnamefont
  {Abdo}} \emph {et~al.} (\bibinfo {collaboration} {Fermi-LAT}),\ }\href
  {\doibase 10.1088/0067-0049/208/2/17} {\bibfield  {journal} {\bibinfo
  {journal} {Astrophys. J. Suppl.}\ }\textbf {\bibinfo {volume} {208}},\
  \bibinfo {pages} {17} (\bibinfo {year} {2013})},\ \Eprint
  {http://arxiv.org/abs/1305.4385} {arXiv:1305.4385 [astro-ph.HE]} \BibitemShut
  {NoStop}%
\bibitem [{\citenamefont {Ackermann}\ \emph {et~al.}(2015)\citenamefont
  {Ackermann} \emph {et~al.}}]{Fermi-LAT:2014ryh}%
  \BibitemOpen
  \bibfield  {author} {\bibinfo {author} {\bibfnamefont {M.}~\bibnamefont
  {Ackermann}} \emph {et~al.} (\bibinfo {collaboration} {Fermi-LAT}),\ }\href
  {\doibase 10.1088/0004-637X/799/1/86} {\bibfield  {journal} {\bibinfo
  {journal} {Astrophys. J.}\ }\textbf {\bibinfo {volume} {799}},\ \bibinfo
  {pages} {86} (\bibinfo {year} {2015})},\ \Eprint
  {http://arxiv.org/abs/1410.3696} {arXiv:1410.3696 [astro-ph.HE]} \BibitemShut
  {NoStop}%
\bibitem [{\citenamefont {Maier}\ \emph {et~al.}(2019)\citenamefont {Maier},
  \citenamefont {Arrabito}, \citenamefont {Bernlöhr}, \citenamefont {Bregeon},
  \citenamefont {Cumani}, \citenamefont {Hassan}, \citenamefont {Hinton},\ and\
  \citenamefont {Moralejo}}]{bookCTA}%
  \BibitemOpen
  \bibfield  {author} {\bibinfo {author} {\bibfnamefont {G.}~\bibnamefont
  {Maier}}, \bibinfo {author} {\bibfnamefont {L.}~\bibnamefont {Arrabito}},
  \bibinfo {author} {\bibfnamefont {K.}~\bibnamefont {Bernlöhr}}, \bibinfo
  {author} {\bibfnamefont {J.}~\bibnamefont {Bregeon}}, \bibinfo {author}
  {\bibfnamefont {P.}~\bibnamefont {Cumani}}, \bibinfo {author} {\bibfnamefont
  {T.}~\bibnamefont {Hassan}}, \bibinfo {author} {\bibfnamefont
  {J.}~\bibnamefont {Hinton}}, \ and\ \bibinfo {author} {\bibfnamefont
  {A.}~\bibnamefont {Moralejo}},\ }\href {\doibase 10.48550/ARXIV.1907.08171}
  {\enquote {\bibinfo {title} {Performance of the cherenkov telescope array},}\
  } (\bibinfo {year} {2019})\BibitemShut {NoStop}%
\bibitem [{\citenamefont {{Maier}}(2019)}]{2019ICRC...36..733M}%
  \BibitemOpen
  \bibfield  {author} {\bibinfo {author} {\bibfnamefont {G.}~\bibnamefont
  {{Maier}}},\ }in\ \href@noop {} {\emph {\bibinfo {booktitle} {36th
  International Cosmic Ray Conference (ICRC2019)}}},\ \bibinfo {series}
  {International Cosmic Ray Conference}, Vol.~\bibinfo {volume} {36}\ (\bibinfo
  {year} {2019})\ p.\ \bibinfo {pages} {733},\ \Eprint
  {http://arxiv.org/abs/1907.08171} {arXiv:1907.08171 [astro-ph.IM]}
  \BibitemShut {NoStop}%
\bibitem [{\citenamefont {Dembinski}\ and\ \citenamefont
  {et~al.}(2020)}]{iminuit}%
  \BibitemOpen
  \bibfield  {author} {\bibinfo {author} {\bibfnamefont {H.}~\bibnamefont
  {Dembinski}}\ and\ \bibinfo {author} {\bibfnamefont {P.~O.}\ \bibnamefont
  {et~al.}},\ }\href {\doibase 10.5281/zenodo.3949207} {\  (\bibinfo {year}
  {2020}),\ 10.5281/zenodo.3949207}\BibitemShut {NoStop}%
\bibitem [{\citenamefont {Buchner}\ \emph {et~al.}(2014)\citenamefont
  {Buchner}, \citenamefont {Georgakakis}, \citenamefont {Nandra}, \citenamefont
  {Hsu}, \citenamefont {Rangel}, \citenamefont {Brightman}, \citenamefont
  {Merloni}, \citenamefont {Salvato}, \citenamefont {Donley},\ and\
  \citenamefont {Kocevski}}]{Buchner:2014nha}%
  \BibitemOpen
  \bibfield  {author} {\bibinfo {author} {\bibfnamefont {J.}~\bibnamefont
  {Buchner}}, \bibinfo {author} {\bibfnamefont {A.}~\bibnamefont
  {Georgakakis}}, \bibinfo {author} {\bibfnamefont {K.}~\bibnamefont {Nandra}},
  \bibinfo {author} {\bibfnamefont {L.}~\bibnamefont {Hsu}}, \bibinfo {author}
  {\bibfnamefont {C.}~\bibnamefont {Rangel}}, \bibinfo {author} {\bibfnamefont
  {M.}~\bibnamefont {Brightman}}, \bibinfo {author} {\bibfnamefont
  {A.}~\bibnamefont {Merloni}}, \bibinfo {author} {\bibfnamefont
  {M.}~\bibnamefont {Salvato}}, \bibinfo {author} {\bibfnamefont
  {J.}~\bibnamefont {Donley}}, \ and\ \bibinfo {author} {\bibfnamefont
  {D.}~\bibnamefont {Kocevski}},\ }\href {\doibase 10.1051/0004-6361/201322971}
  {\bibfield  {journal} {\bibinfo  {journal} {Astron. Astrophys.}\ }\textbf
  {\bibinfo {volume} {564}},\ \bibinfo {pages} {A125} (\bibinfo {year}
  {2014})},\ \Eprint {http://arxiv.org/abs/1402.0004} {arXiv:1402.0004
  [astro-ph.HE]} \BibitemShut {NoStop}%
\bibitem [{\citenamefont {Abramowski}\ \emph {et~al.}(2016)\citenamefont
  {Abramowski} \emph {et~al.}}]{HESS:2016pst}%
  \BibitemOpen
  \bibfield  {author} {\bibinfo {author} {\bibfnamefont {A.}~\bibnamefont
  {Abramowski}} \emph {et~al.} (\bibinfo {collaboration} {H.E.S.S.}),\ }\href
  {\doibase 10.1038/nature17147} {\bibfield  {journal} {\bibinfo  {journal}
  {Nature}\ }\textbf {\bibinfo {volume} {531}},\ \bibinfo {pages} {476}
  (\bibinfo {year} {2016})},\ \Eprint {http://arxiv.org/abs/1603.07730}
  {arXiv:1603.07730 [astro-ph.HE]} \BibitemShut {NoStop}%
\bibitem [{\citenamefont {Hooper}\ \emph {et~al.}(2018)\citenamefont {Hooper},
  \citenamefont {Cholis},\ and\ \citenamefont {Linden}}]{Hooper:2017rzt}%
  \BibitemOpen
  \bibfield  {author} {\bibinfo {author} {\bibfnamefont {D.}~\bibnamefont
  {Hooper}}, \bibinfo {author} {\bibfnamefont {I.}~\bibnamefont {Cholis}}, \
  and\ \bibinfo {author} {\bibfnamefont {T.}~\bibnamefont {Linden}},\ }\href
  {\doibase 10.1016/j.dark.2018.05.004} {\bibfield  {journal} {\bibinfo
  {journal} {Phys. Dark Univ.}\ }\textbf {\bibinfo {volume} {21}},\ \bibinfo
  {pages} {40} (\bibinfo {year} {2018})},\ \Eprint
  {http://arxiv.org/abs/1705.09293} {arXiv:1705.09293 [astro-ph.HE]}
  \BibitemShut {NoStop}%
\bibitem [{\citenamefont {Evoli}\ \emph {et~al.}(2018)\citenamefont {Evoli},
  \citenamefont {Linden},\ and\ \citenamefont {Morlino}}]{Evoli:2018aza}%
  \BibitemOpen
  \bibfield  {author} {\bibinfo {author} {\bibfnamefont {C.}~\bibnamefont
  {Evoli}}, \bibinfo {author} {\bibfnamefont {T.}~\bibnamefont {Linden}}, \
  and\ \bibinfo {author} {\bibfnamefont {G.}~\bibnamefont {Morlino}},\ }\href
  {\doibase 10.1103/PhysRevD.98.063017} {\bibfield  {journal} {\bibinfo
  {journal} {Phys. Rev. D}\ }\textbf {\bibinfo {volume} {98}},\ \bibinfo
  {pages} {063017} (\bibinfo {year} {2018})},\ \Eprint
  {http://arxiv.org/abs/1807.09263} {arXiv:1807.09263 [astro-ph.HE]}
  \BibitemShut {NoStop}%
\bibitem [{\citenamefont {Fang}\ \emph {et~al.}(2019)\citenamefont {Fang},
  \citenamefont {Bi},\ and\ \citenamefont {Yin}}]{Fang:2019iym}%
  \BibitemOpen
  \bibfield  {author} {\bibinfo {author} {\bibfnamefont {K.}~\bibnamefont
  {Fang}}, \bibinfo {author} {\bibfnamefont {X.-J.}\ \bibnamefont {Bi}}, \ and\
  \bibinfo {author} {\bibfnamefont {P.-F.}\ \bibnamefont {Yin}},\ }\href
  {\doibase 10.1093/mnras/stz1974} {\bibfield  {journal} {\bibinfo  {journal}
  {Mon. Not. Roy. Astron. Soc.}\ }\textbf {\bibinfo {volume} {488}},\ \bibinfo
  {pages} {4074} (\bibinfo {year} {2019})},\ \Eprint
  {http://arxiv.org/abs/1903.06421} {arXiv:1903.06421 [astro-ph.HE]}
  \BibitemShut {NoStop}%
\bibitem [{\citenamefont {Crocker}\ \emph {et~al.}(2010)\citenamefont
  {Crocker}, \citenamefont {Jones}, \citenamefont {Melia}, \citenamefont
  {Ott},\ and\ \citenamefont {Protheroe}}]{Crocker:2010xc}%
  \BibitemOpen
  \bibfield  {author} {\bibinfo {author} {\bibfnamefont {R.~M.}\ \bibnamefont
  {Crocker}}, \bibinfo {author} {\bibfnamefont {D.}~\bibnamefont {Jones}},
  \bibinfo {author} {\bibfnamefont {F.}~\bibnamefont {Melia}}, \bibinfo
  {author} {\bibfnamefont {J.}~\bibnamefont {Ott}}, \ and\ \bibinfo {author}
  {\bibfnamefont {R.~J.}\ \bibnamefont {Protheroe}},\ }\href {\doibase
  10.1038/nature08635} {\bibfield  {journal} {\bibinfo  {journal} {Nature}\
  }\textbf {\bibinfo {volume} {468}},\ \bibinfo {pages} {65} (\bibinfo {year}
  {2010})},\ \Eprint {http://arxiv.org/abs/1001.1275} {arXiv:1001.1275
  [astro-ph.GA]} \BibitemShut {NoStop}%
\bibitem [{\citenamefont {Macias}\ \emph {et~al.}(2021)\citenamefont {Macias},
  \citenamefont {van Leijen}, \citenamefont {Song}, \citenamefont {Ando},
  \citenamefont {Horiuchi},\ and\ \citenamefont {Crocker}}]{Macias:2021boz}%
  \BibitemOpen
  \bibfield  {author} {\bibinfo {author} {\bibfnamefont {O.}~\bibnamefont
  {Macias}}, \bibinfo {author} {\bibfnamefont {H.}~\bibnamefont {van Leijen}},
  \bibinfo {author} {\bibfnamefont {D.}~\bibnamefont {Song}}, \bibinfo {author}
  {\bibfnamefont {S.}~\bibnamefont {Ando}}, \bibinfo {author} {\bibfnamefont
  {S.}~\bibnamefont {Horiuchi}}, \ and\ \bibinfo {author} {\bibfnamefont
  {R.~M.}\ \bibnamefont {Crocker}},\ }\href {\doibase 10.1093/mnras/stab1450}
  {\bibfield  {journal} {\bibinfo  {journal} {Mon. Not. Roy. Astron. Soc.}\
  }\textbf {\bibinfo {volume} {506}},\ \bibinfo {pages} {1741} (\bibinfo {year}
  {2021})},\ \Eprint {http://arxiv.org/abs/2102.05648} {arXiv:2102.05648
  [astro-ph.HE]} \BibitemShut {NoStop}%
\end{thebibliography}%

\end{document}